\documentclass[12pt,draftcls,onecolumn] {IEEEtran}

\usepackage[T1]{fontenc}
\ifCLASSINFOpdf
\else
\fi
\usepackage{array}
\usepackage{multirow}
\usepackage{cite}
\usepackage{amsmath}
\usepackage{booktabs}
\usepackage{color}
\usepackage{graphicx}
\usepackage{mathrsfs}
\usepackage{tikz}
\usepackage{calc}
\usepackage{amssymb}
\usepackage[cmintegrals]{newtxmath}
\usepackage{threeparttable}

\begin{document}
\renewcommand\arraystretch{1}

\title{Low-Complexity Improved-Throughput Generalised Spatial Modulation: Bit-to-Symbol Mapping, Detection and Performance Analysis}

\author{Jiancheng~An, Chao~Xu, \emph{Senior Member, IEEE}, Yusha~Liu, Lu~Gan, and~Lajos~Hanzo, \emph{Fellow, IEEE}
	\thanks{This work was supported in part by the China Scholarship Council. The work of L. Hanzo was supported in part by the Engineering and Physical Sciences Research Council projects EP/Noo4558/1, EP/PO34284/1, and COALESCE, in part by the Royal Society’s Global Challenges Research Fund Grant, and in part by the European Research Council’s Advanced Fellow Grant QuantCom.}
	\thanks{J. An and L. Gan are with the School of Information and Communication Engineering, University of Electronic Science and Technology of China, Chengdu, Sichuan 611731, China. (E-mail: jiancheng$\_$an@163.com; ganlu@uestc.edu.cn).}
	\thanks{C. Xu, Y. Liu and L. Hanzo are with the School of Electronics and Computer Science, University of Southampton, SO17 1BJ, UK. (E-mail: cx1g08@soton.ac.uk; yl6g15@ecs.soton.ac.uk; lh@ecs.soton.ac.uk).}
	}


\maketitle

\begin{abstract}
Low-complexity improved-throughput generalised spatial modulation (LCIT-GSM) is proposed. More explicitly, in GSM, extra information bits are conveyed implicitly by activating a fixed number $N_{a}$ out of $N_{t}$ transmit antennas (TAs) at a time. As a result, GSM has the advantage of a reduced number of radio-frequency (RF) chains and reduced inter-antenna interference (IAI) at the cost of a lower throughput than its multiplexing-oriented full-RF based counterparts. Variable-${N_a}$ GSM mitigates this throughput reduction by incorporating all possible TA activation patterns associated with a variable value $N_{a}$ ranging from $1$ to $N_{t}$ during a single channel-use, which maximises the throughput of GSM but suffers a high complexity of the mapping book design and demodulation. In order to mitigate the complexity, \emph{first of all}, we propose two efficient schemes for mapping the information bits to the TA activation patterns, which can be readily scaled to massive MIMO setups. \emph{Secondly}, in the absence of IAI, we derive a pair of low-complexity near-optimal detectors, one of them has a reduced search scope, while the other benefits from a decoupled single-stream based signal detection algorithm. \emph{Finally}, the performance of the proposed LCIT-GSM system is characterised by the error probability upper bound (UB). Our Monte Carlo based simulation results confirm the improved error performance of our proposed scheme, despite its reduced signal detection complexity.
\end{abstract}


\IEEEpeerreviewmaketitle

\section{Introduction}\label{s1}
Spatial modulation (SM), constitutes a promising technique for the next-generation multiple-input multiple-output (MIMO) system, which has attracted substantial research interests \cite{mesleh2006spatial, mesleh2008spatial, di2013spatial}. The SM scheme strikes a flexible trade-off between conventional MIMO techniques including Alamouti's space-time block code (STBC), and vertical Bell Laboratories layered space-time (V-BLAST) system \cite{foschini1996layered, tarokh1999space, zheng2003diversity}, which aim for retaining a diversity gain and a multiplexing gain, respectively. More explicitly, for the SM scheme, the indices of the transmit antenna (TA) are exploited as an extra dimension for transmitting information besides the conventional amplitude-phase shift keying (APSK) such as phase-shift keying (PSK) and quadrature amplitude modulation (QAM) \cite{di2011spatial, di2013spatial, yang2014design}. Owing to the fact that only a single TA is activated in SM-MIMO at any time instant, the inter-channel interference (ICI) and the inter-antenna interference (IAI) of conventional MIMO techniques are mitigated, which results in the simplified transceiver design and a reduced signal detection complexity \cite{mesleh2008spatial, di2013spatial}. Specifically, the MIMO power consumption is effectively reduced by the SM philosophy, where only a single power amplifier is required \cite{sugiura2014single, masouros2015constellation, xu2019sixty}. As a further advance, the SM scheme can be flexibly configured for challenging communication scenarios, such as that of the downlink MIMO systems having a rank-deficient channel matrix \cite{yang2014design}.

Inspired by the aforementioned advantages, the space shift keying (SSK) concept was introduced in \cite{jeganathan2008generalized, jeganathan2009space} as a low-complexity implementation of SM, where only the TA indices convey information. As a further advance, the principle of SM has also been extended to the time domain, resulting in the space-time shift keying (STSK) \cite{sugiura2010coherent, xu2019near, sugiura2011generalized, xu2019constant}, to the frequency domain yielding the space-frequency shift keying (SFSK) \cite{li2018space} and to the time-frequency domain creating the space-time-frequency shift keying (STFSK) \cite{ngo2011space, ngo2012area} to exploit all three degrees of freedom. As a result, the generalised concept of index modulation (IM), has gained widespread interest in both academia and industry \cite{basar2016index, basar2017index, lu2018compressed, mao2018novel}. Furthermore, in \cite{mesleh2010indoor, mesleh2011optical} and \cite{kabir2015d}, the IM philosophy was also extended to indoor optical wireless communication and molecular communication, where visible light and molecules are the medium of information transmission, respectively. Besides, both the analytical studies, numerical simulations, as well as real-world experiments have verified that SM-MIMOs have the inherent potential of outperforming many state-of-art MIMO schemes, provided that a sufficiently high number of TAs is available at the transmitter \cite{serafimovski2013practical, younis2013performance}.

\begin{table*}[!t]
\renewcommand\arraystretch{1.5}
\centering
	\caption{\label{tab1} The contribution of the proposed LCIT-GSM scheme compared with other GSM schemes.}
	\begin{threeparttable}
\begin{tabular}{|l||c||c|c|c|c|c|c|c|c|c|c|}
\hline
\textbf{Contributions} &  \textbf{LCIT-GSM} & \textbf{\cite{xiao2018single}} & \textbf{\cite{koundinya2016joint}}& \textbf{\cite{he2017spatial}} & \textbf{\cite{an2016mutual}} & \textbf{\cite{xiao2019compressive}}& \textbf{\cite{guo2019signal}}& \textbf{\cite{kabir2015d}}& \textbf{\cite{osman2015variable}}& \textbf{\cite{wang2012generalised}}& \textbf{\cite{liu2017variable}} \\ \hline
\textbf{Flexibility in the number of TAs} &  \checkmark & \checkmark & $\times$ & \checkmark & \checkmark & \checkmark & \checkmark& \checkmark& \checkmark& \checkmark& \checkmark  \\ \hline
\textbf{Multiple active TAs} &  \checkmark & \checkmark & $\times$ & \checkmark & \checkmark & \checkmark &  \checkmark& \checkmark& \checkmark & \checkmark& \checkmark \\ \hline
\textbf{Multiple streams$^ * $} &  $\times$ & \checkmark & $\times$ & \checkmark & \checkmark & \checkmark &  \checkmark& $\times$& $\times$ & \checkmark& $\times$ \\ \hline
\textbf{Variable number of active TAs}  & \checkmark & $\times$ & $\times$ & $\times$ & $\times$ & $\times$  &$\times$& \checkmark& \checkmark&$\times$& \checkmark\\ \hline
\textbf{Bitwise spatial mapping}  & \checkmark & $\times$ & $\times$ & $\times$ & $\times$ & $\times$ &$\times$& \checkmark&$\times$&$\times$&$\times$ \\ \hline
\textbf{Joint spatial and classic symbol alphabet}  & \checkmark & \checkmark & \checkmark & $\times$ & \checkmark & $\times$ & \checkmark& $\times$&$\times$& \checkmark &$\times$\\ \hline
\textbf{Low-complexity detectors} &  \checkmark & \checkmark & $\times$ & $\times$ & $\times$ & $\times$ &$\times$&$\times$ &$\times$ & \checkmark&$\times$ \\ \hline
\textbf{Improved upper bound} & \checkmark & $\times$ & \checkmark & $\times$ &  $\times$ & $\times$ &$\times$&$\times$&$\times$ &$\times$&$\times$\\ \hline
\end{tabular}
\begin{tablenotes}\footnotesize
\item \textbf{\cite{xiao2018single}-2018}: Single-RF and twin-RF SM for an arbitrary number of TAs;
\item \textbf{\cite{koundinya2016joint}-2016}: Joint spatial and classic symbol alphabet for SM;
\item \textbf{\cite{he2017spatial}-2017}: GSM aided mm-wave MIMO;
\item \textbf{\cite{an2016mutual}-2016}: Diagonal precoding aided GSM;
\item \textbf{\cite{xiao2019compressive}-2019}: Generalized quadrature spatial modulation;
\item \textbf{\cite{guo2019signal}-2019}: Generic signal shaping methods for the multiple-stream GSM;
\item \textbf{\cite{kabir2015d}-2015}: Depleted molecule shift keying;
\item \textbf{\cite{osman2015variable}-2015}: Variable active antenna SM;
\item \textbf{\cite{wang2012generalised}-2012}: GSM with multiple active TAs;
\item \textbf{\cite{liu2017variable}-2017}: Variable-$N_{a}$ GSM.
    \end{tablenotes}
    $^ * $The LCIT-GSM scheme considered in this paper is a single-stream GSM scheme that improves the throughput by using more TA activation patterns in the spatial domain while inherits the advantage of IAI-free. We note that the LCIT-GSM scheme can also be combined with V-BLAST to improve throughput to a greater extent, which, however, results in a variable-length code with serious error propagation.
      \end{threeparttable}\\
\dotfill
\end{table*}

In order to improve the throughput of single-RF SM and SSK schemes, generalised spatial modulation (GSM) was proposed in \cite{younis2010generalised, fu2010generalised}, which activates several TAs simultaneously. Furthermore, conventional full-RF diversity- and multiplexing-oriented MIMO schemes have been combined with GSM \cite{basar2010space, wang2012generalised, datta2015generalized, xiao2017compressed}. Specifically, in \cite{basar2010space}, the space-time block coding-spatial modulation (STBC-SM) concept was proposed, where GSM was combined with STBC in order to achieve a beneficial diversity gain at the cost of using a reduced number of RF chains. Similarly, a novel multiplexing-oriented GSM scheme was proposed in \cite{wang2012generalised, datta2015generalized, xiao2017compressed} that combines V-BLAST and GSM, where multiple TAs are activated simultaneously in order to transmit independently modulated symbols. This scheme imposed a reduced IAI compared to the full-RF V-BLAST scheme. Recently, a new transmission scheme termed as generalised quadrature spatial modulation (GQSM) was proposed in \cite{xiao2019compressive}, which extends the GSM constellation to the in-phase and quadrature dimensions \cite{mesleh2014quadrature}. The QSM mapping of \cite{mesleh2014quadrature} has then been further improved in \cite{li2016generalized, mohaisen2018increasing, mesleh2017transmitter, mesleh2018generalized} by increasing the system's throughput and/or reducing the bit error probability (BEP). However, all the aforementioned GSM schemes -- including the STBC-SM, SM+V-BLAST and GQSM -- always employ a fixed number $N_{a}$ of TAs, which inevitably results in the well-known throughput limitation of the family of GSM schemes. In order to eliminate this throughput limitation, the variable-$N_{a}$ GSM (VGSM) concept was proposed for increasing the spatial constellation size by employing a variable number $N_{a}$ of activated TAs \cite{osman2015variable, younis2014spatial, mesleh2018space}.

\begin{figure}[!t]
	\centering
	\includegraphics[width=8.8cm]{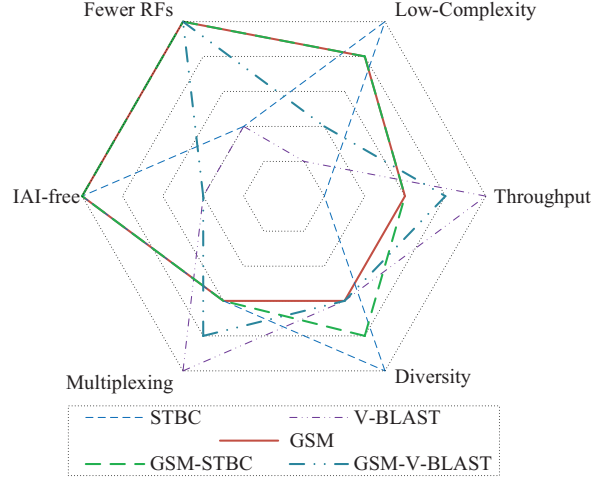}
	\caption{``Spider diagram" comparing several space-time transmission schemes. The hierarchy structure models ordered MIMO schemes in each category. From the outer to the inner, the gain of the corresponding term becomes smaller for a selected direction. Taking the throughput as an example, V-BLAST has the highest throughput, followed by GSM+V-BLAST, GSM, GSM+STBC, SM and STBC.}
	\label{f1}
\end{figure}

Against this background, we propose a low-complexity improved-throughput generalised spatial modulation (LCIT-GSM) scheme, which is inspired by the novel bit-to-symbol mapping scheme of \cite{kabir2015d} in the context of molecular communication. In order to reflect on these issues, first of all, the trade-offs between the aforementioned state-of-the-art MIMO schemes are visualised by the spider diagram of Fig. \ref{f1}. The rationale of Fig. \ref{f1} is that considering the throughput as an example, V-BLAST achieves the highest throughput, followed by GSM+V-BLAST, GSM, GSM+STBC and STBC. Therefore, there are five rings in Fig. \ref{f1}, which represent the ordered position of the five main MIMO families in each category. As summarised in Fig. \ref{f1}, owing to its fixed number of activated TAs, the GSM schemes generally suffer from a throughput loss. Furthermore, the features of the new LCIT-GSM scheme using the improved bitwise spatial mapping design of \cite{kabir2015d} are compared to the existing GSM schemes in Table \ref{tab1}, where the specific terminologies of the systems referenced are listed below Table \ref{tab1} for the readers' convenience. More explicitly, the contributions of the paper are as follows:

\begin{figure*}[!t]
	\centering
	\includegraphics[width=16.5cm]{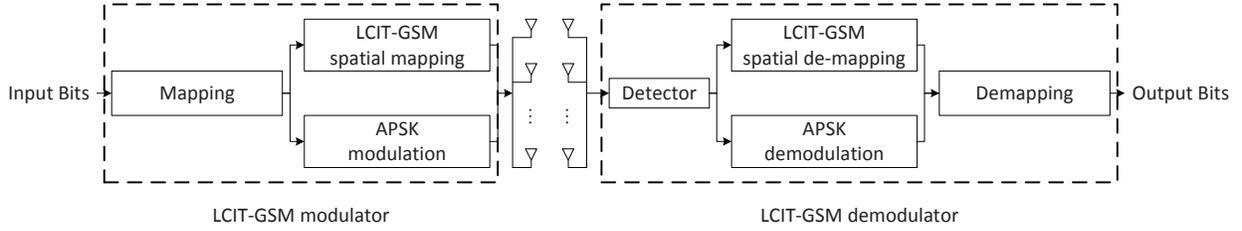}
	\caption{Schematic of the LCIT-GSM system model, where the number of activated TAs is a variable that carries source information.}
	\label{f2}
\end{figure*}
\begin{itemize}
\item First of all, a LCIT-GSM scheme is proposed for improving the throughput by activating a variable number $N_{a}$ of activated TAs, while reducing the complexity of the mapping design. More specifically, we introduce three spatial mapping arrangements that incorporate all the legitimate TA activation patterns. The first two of them are motivated by the so-called depleted molecule shift keying (MoSK) scheme of \cite{kabir2015d}. In order to accommodate the special case of all-zeros in the bitstream that may result in no TA being activated and hence no modulated symbols being transmitted, we conceive specifically designed mapping arrangements for both SM and the classic APSK modulation. This is in contrast to the MoSK regime of \cite{kabir2015d}, where the all-zero state of the nano-transmitter is directly exploited to convey information bits in the context of SSK-based molecular communication. Furthermore, a look-up table (LUT) based spatial mapping scheme similar to \cite{osman2015variable, younis2014spatial} is also included in order to demonstrate the advantages of the proposed low-complexity detectors.

\item Secondly, in the absence of IAI in the LCIT-GSM mapping, we propose a pair of low-complexity detectors based on the classical maximum likelihood detector (MLD). Explicitly, a two-step near-maximum likelihood detector (TMLD) is derived, which substantially reduces the size of the legitimate set to be searched. Furthermore, a decoupling based maximum likelihood detector (DMLD) is conceived for decoupling a two-dimensional joint search into a pair of one-dimensional single-stream based searches, while maintaining the optimal detection performance.

\item Thirdly, based on the prior work of \emph{Haas et al.} \cite{di2012bit}, we derive a tighter performance upper bound (UB) for the proposed LCIT-GSM system. Explicitly, the improved performance bound is composed of three parts, which respectively reflect the impacts of the LCIT-GSM scheme's signal constellation diagram, spatial constellation diagram and their interactions. Finally, our Monte Carlo based simulation results confirm the improved spectral efficiency (SE), reduced signal detection complexity and the improved BEP performance of our proposed LCIT-GSM scheme compared to conventional GSM scheme \cite{younis2010generalised} and VGSM scheme \cite{osman2015variable, younis2014spatial}.
\end{itemize}

The rest of this paper is structured as follows. Section \ref{s2} presents the improved LCIT-GSM system model and spatial mapping schemes. Following this, our detection algorithms are conceived in Section \ref{s3}. Section \ref{s4} analyses the theoretical BEP and computational complexity of the LCIT-GSM system, while the BEP performance of our LCIT-GSM systems is presented in Section \ref{s5}. Finally, our conclusions are offered in Section \ref{s6}.

\emph{Notation:} We use upper (lower) bold face letters to indicate matrices (column vectors). ${\left(  \cdot  \right)^T}$, ${\left(  \cdot  \right)^H}$ and ${\left(  \cdot  \right)^ * }$ represent the transpose, Hermitian transpose and conjugate, respectively. $E\left\{  \cdot  \right\}$ stands for the expected value. $\left\|  \cdot  \right\|$ is the Frobenius norm of a complex vector. $\otimes $ represents the Kronecker product. We denote the $N \times N$ identity matrix as ${{\bf{I}}_N}$. ${\bf{0}}$ and ${\bf{1}}$ denote an all-zero vector and an all-one vector, respectively, with appropriate dimensions. Furthermore, $\left\lfloor  \cdot  \right\rfloor$ and $\left \lfloor \cdot  \right \rceil$ represent the floor and rounding operation, respectively. The $\log \left(  \cdot  \right)$ functions are base-2 by default. $C_n^k$ denotes the number of combinations of taking $k$ out of $n$. ${{\mathcal{M}}_\gamma }\left( t \right) = E\left\{ {\exp \left( { - \gamma t} \right)} \right\}$ is the moment generating function (MGF) of the random variable $\gamma $. $Q\left(x\right)=\frac{1}{\sqrt{2\pi }}{\int }_{x}^{+\infty }exp\left(\frac{-{t}^{2}}{2}\right)dt$ is the Q-function. $\angle \left ( z \right )$ denotes the phase angle of a complex number $z$. $\Re \left ( z \right )$ and $\Im \left ( z \right )$ denote the real and imaginary part of a complex number $z$, respectively.

\section{System Model}\label{s2}

\subsection{LCIT-GSM Transmission}
The LCIT-GSM system model is depicted in Fig. \ref{f2}. The incoming data bits are first grouped into $m = {m_s} + {m_a}$ bits. Following this, the grouped block of bits ${{\bf{b}}^T}$ is split into $m_s$ spatial bits ${\bf{b}}_s^T$ and $m_a$ APSK modulated symbol bits ${\bf{b}}_a^T$, i.e., ${{\bf{b}}^T} = \left[ {\left. {{\bf{b}}_s^T} \right|{\bf{b}}_a^T} \right]$. The first $m_s$ spatial bits are assigned to select the TA activation pattern following the spatial mapping procedures, which will be detailed in Section \ref{2_2}. We note that only a reduced number of TAs are activated at any time. The remaining $m_a$ bits are modulated using APSK, such as $M$-PSK or $M$-QAM, where the modulation order is given by $M = {2^{{m_a}}}$.

In contrast to the classic GSM, which activates a fixed number of TAs at each slot, the number of activated TAs in the LCIT-GSM scheme is variable \cite{osman2015variable, younis2014spatial}. This constitutes the most distinguished benefit of the LCIT-GSM design, where the maximum throughput is pursued by utilizing all legitimate TA activation patterns. The total number $N_m$ of all legitimate TA activation patterns is given by:
\begin{equation}\label{eq2-1-1}
{N_m} = \mathop {C_{{N_t}}^0}\limits_{ -  -  -  - }  + C_{{N_t}}^1 + C_{{N_t}}^2 +  \cdots  + C_{{N_t}}^{{N_t}} = 2^{N_{t}},
\end{equation}
where the underlined $C_{{N_t}}^0$ corresponds to the spatial bit sequence having all zeros.

\begin{figure}[!t]
	\centering
	\includegraphics[width=8.8cm]{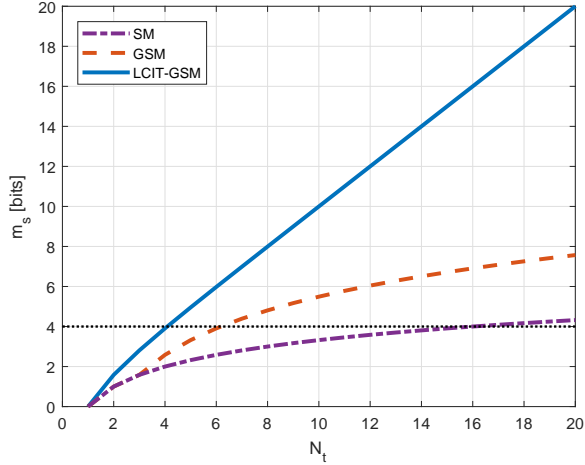}
	\caption{The number $m_s$ of spatial bits versus the number $N_t$ of TAs for SM, GSM, and LCIT-GSM, where we have a fixed number ${N_a} = 2$ of activated TAs for GSM.}
	\label{f3}
\end{figure}

However, it should be noted that the spatial bitstream with all zeros in LCIT-GSM, i.e., ${\bf{b}}_s^T = {\bf{0}}$ cannot transmit an APSK symbol carrying $m_{a}$ bits, hence it must be avoided. This problem will be addressed in Section \ref{2_2}. If the case corresponding to ${\bf{b}}_s^T = {\bf{0}}$ is eliminated, only $\left ( {{N_m} - 1} \right )$ TA activation patterns remain legitimate. Bearing in mind that the number of TA activation patterns that convey $m_{s}$ source bits must be a power of two, the maximum value of ${m_s}$ becomes\footnote{We note that $m_{s}$ is increased beyond the limit of (\ref{eq2-1-2}) by the first of our spatial mapping schemes proposed in Section \ref{2_2}, namely in DTAA-R, where we have $m_{s}=\log\left ( 2^{N_{t}} \right )=N_{t}$. The absence of the all-zero activation pattern is compensated by the transformation of the signal constellation, which ensures maintaining the maximum transmission rate combined with the low complexity of the direct bit-to-antenna mapping scheme. In this subsection, we assume that $\left( {{N_t} - 1} \right)$ spatial bits are conveyed.}

\begin{equation}\label{eq2-1-2}
{m_s} = \left\lfloor {\log \left( {{N_m} - 1} \right)} \right\rfloor  = \left\lfloor {\log \left( {2^{N_{t}} - 1} \right)} \right\rfloor  = {N_t} - 1.
\end{equation}

In summary, the total number of source bits conveyed by the LCIT-GSM design is given by
\begin{equation}\label{eq2-1-3}
m = {m_a} + {m_s} = \log M + {N_t} - 1,
\end{equation}
which is evidently higher than that of GSM given by $\log M + \left \lfloor \log C_{{N_t}}^{{N_a}} \right \rfloor$, where $N_a$ is the number of RF chains. Fig. \ref{f3} compares the number $m_s$ of spatial bits versus the number $N_t$ of TAs, for using GSM and LCIT-GSM, where the case of SM is portrayed as a special case of GSM associated with ${N_a} = 1$. It can be seen that when $m_s = 4$ spatial bits are transmitted, $N_t = 4$ is needed for LCIT-GSM. For the same throughput, the GSM scheme would require an increased value of $N_{t}=6$, while the SM scheme requires a further increased value of $N_t = 16$, as demonstrated by Fig. \ref{f3}.

\emph{Remark 1:} We note that, first of all, the number of spatial bits in LCIT-GSM increases with the number of RF chains. By contrast, in GSM, the number of spatial bits decreases when $N_a$ is increased beyond $\left \lfloor \frac{N_{t}}{2} \right \rfloor$ \cite{younis2010generalised}. Although we directly introduce LCIT-GSM as having $N_t$ RF chains, hence exhibiting the maximum throughput improvement, it must be mentioned that our system is also eminently suitable for a limited number of RF chains. Secondly, we note that the same APSK symbol is transmitted by all activated TAs in order to retain the IAI-free feature. Finally, similar to the GSM scheme's combination with STBC and V-BLAST \cite{xiao2018bayesian, xiao2017compressed}, the proposed LCIT-GSM can also be combined with the classic MIMO schemes for the sake of striking further trade-offs between throughput, complexity and diversity, which is left for future work.

In the LCIT-GSM system, the transmit signal is modelled by a ${N_t} \times 1$ column vector $\bf{x}$, which is transmitted over an ${N_t} \times {N_r}$ MIMO Rayleigh flat fading wireless channel, ${\bf{H}} = [ {{{\bf{h}}_1}}\quad{{{\bf{h}}_2}}\quad \cdots \quad{{{\bf{h}}_{{N_t}}}}]$, where ${{\bf{h}}_{k'}}$ is the $k'$-th column of $\bf{H}$ that represents the channel vector of the link spanning from the $k'$-th TA to all of $N_{r}$ receive antennas (RAs). The entries of ${\bf{H}}$ are generated by complex independent and identically distributed (i.i.d.) Gaussian random variables having zero-mean and unit-variance.
\begin{figure*}[!t]
	\centering
	\includegraphics[width=16.5cm]{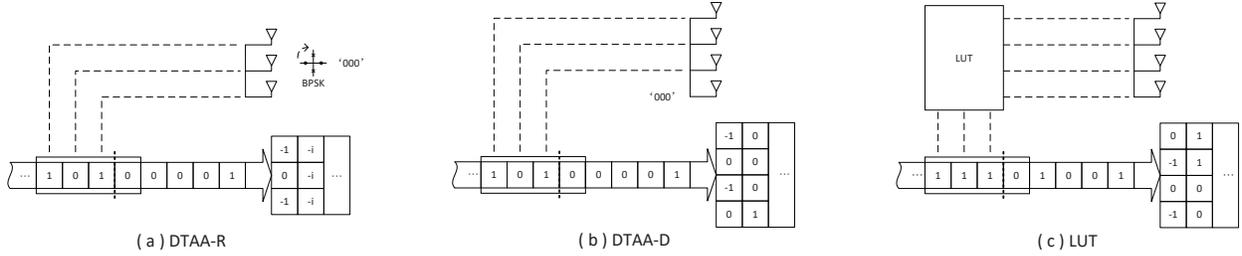}
	\caption{(a) DTAA-R mapping scheme: The $i$-th spatial bit controls the on/off state of the $i$-th TA, where $1\leq i\leq 3$. For the special case of '000', all the three TAs are activated, while the $M$-PSK/QAM symbol is rotated according to (\ref{eq2-2-2}) or (\ref{eq2-2-3}); (b) DTAA-D mapping scheme: The $i$-th spatial bit controls the on/off state of the $i$-th TA, where $1\leq i\leq 3$. The 4-th TA is only activated when the spatial bitstream of '000' is encountered; (c) LUT mapping scheme: The combination book selects the active TAs based on the current spatial bitstream, which uses a mapping design similar to the VGSM of \cite{osman2015variable, younis2014spatial}.}
	\label{f4}
\end{figure*}

Finally, the signal ${\bf{y}}$ received at any symbol instant is given by
\begin{equation}\label{eq2-1-4}
{\bf{y}} = {\bf{Hx}} + {\bf{n}} = {{\bf{h}}_{{{\mathcal {S}}_k}}}s + {\bf{n}},
\end{equation}
where $s \in M$-APSK is the modulated $M$-PSK/QAM symbol\footnote{We note that in the first of the proposed bit-to-symbol mapping schemes detailed in Section \ref{2_2}, namely in DTAA-R, $s$ represents the rotated PSK/QAM symbol when the all-zero spatial bitstream is encountered.}, while $\mathcal{S}_k$ is the $k$-th TA activation combination associated with $k \in \left\{ {1,2, \cdots ,{2^{N_t-1}}} \right\}$. The mapping mechanism of $\left\{ {{\mathcal{S}_1},{\mathcal{S}_2}, \cdots ,{\mathcal{S}_{{2^{N_t-1}}}}} \right\}$ will be detailed in Section \ref{2_2}. Furthermore, the channel vector can be expressed as
\begin{equation}\label{eq2-1-5}
    {{\bf{h}}_{{\mathcal{S}_k}}} = \sum\limits_{k \in {\mathcal{S}_k}} {{{\bf{h}}_k}},
\end{equation}
which is the summation of all channel vectors corresponding to the $k$-th TA activation patterns. Moreover, ${\bf{n}}$ in (\ref{eq2-1-4}) is an AWGN vector associated with zero-mean and variance of $\sigma _n^2$ in each dimension at the receiver.
\subsection{LCIT-GSM spatial mapping scheme}\label{2_2}
In this subsection, we will introduce three spatial mapping arrangements for LCIT-GSM systems, which are: \emph{1)} direct TA activation relying on constellation rotation (DTAA-R), \emph{2)} direct TA activation relying on a dedicated TA (DTAA-D) and \emph{3)} LUT-based spatial mapping (LUT). More specifically, DTAA-R and DTAA-D are the improved schemes of the bitwise activation scheme of the MoSK \cite{kabir2015d}, while LUT uses the same mapping design as VGSM of \cite{osman2015variable, younis2014spatial}, which we include in order to share the low-complexity detectors proposed in Section \ref{s3}.

\textbf{1) DTAA-R:}
The motivation of the DTAA-R scheme is to establish a direct mapping between the spatial bits and the TA activation pattern, which does not require a large LUT. The simplest way of implementing this is to assign each spatial bit to independently control the on/off state of the TAs, as portrayed in Fig. 4 (a). Explicitly, the spatial bit-pattern `101' activates the upper- and lower-most TAs and the `0' bit in the fourth position of the box is BPSK-modulated onto both active TAs. The TA in the middle is deactivated. However, as mentioned before, when all spatial bits of $b_{s}$ are zeros, no TAs are activated to transmit the classic $M$-PSK/QAM symbol, which constitutes a challenge in our DTAA-R design. In order to solve this problem, we propose the following revised DTAA-R mapping rule:
\begin{equation}\label{eq2-2-1}
\left\{ {\begin{array}{*{20}{l}}
{{{\bf{b}}_s}^T \ne {\bf{0}} \Leftrightarrow {\bf{x}} = {\bf{b}}_s \times s}\\
{{{\bf{b}}_s}^T = {\bf{0}} \Leftrightarrow {\bf{x}} = {\bf{1}} \times T\left( s \right)},
\end{array}} \right.
\end{equation}
where ${\bf{1}}$ is a ${N_t} \times 1$ column vector of ones. In this way, all TAs are activated, when the spatial bits are all zeros. Moreover, owing to the fact that the TA activation pattern corresponding to $\mathbf{b}_{s}^{T}=\mathbf{0}$ becomes exactly the same as in the case of $\mathbf{b}_{s}^{T}=\mathbf{1}$, we propose to apply a transformation $T\left(  \cdot  \right)$ to the APSK constellation.

More explicitly, for an $M$-PSK symbol, a simple phase rotation is applied to the APSK symbol, when all the TAs are activated for $\mathbf{b}_{s}^{T}=\mathbf{0}$, which can be expressed as:
\begin{equation}\label{eq2-2-2}
T_{PSK}\left( s \right) = {e^{j\frac{\pi }{M}}} \cdot s.
\end{equation}

\begin{table*}[!t]
\renewcommand\arraystretch{1.5}
    	\centering
	\caption{\label{tab2} LCIT-GSM mapping table for DTAA-R, DTAA-D, LUT, respectively, where $\left( { \cdot , \cdot } \right)$ indicates the indices of the active TAs, while $N_{t}=\left \{ 3,4 \right \}$, $m_{s}=3$ and $m_{a}=1$ are assumed$^*$.}
	\begin{threeparttable}
\begin{tabular}{|c|c|c|c|c|c|c|}
\hline
\multirow{2}{*}{Input Bits} & \multicolumn{2}{c|}{DTAA-R with ${N_t} = 3$} & \multicolumn{2}{c|}{DTAA-D with ${N_t} = 4$} & \multicolumn{2}{c|}{LUT with ${N_t} = 4$} \\ \cline{2-7} 
 & TA Activation Pattern & Symbol & TA Activation Pattern & Symbol & TA Activation Pattern & Symbol \\ \hline
0000 & (1,2,3) & \emph{i} & (4) & -1 & (1) & -1 \\ \hline
0001 & (1,2,3) & \emph{-i} & (4) & 1 & (1) & 1 \\ \hline
0010 & (3) & -1 & (3) & -1 & (2) & -1 \\ \hline
0011 & (3) & 1 & (3) & 1 & (2) & 1 \\ \hline
0100 & (2) & -1 & (2) & -1 & (3) & -1 \\ \hline
0101 & (2) & 1 & (2) & 1 & (3) & 1 \\ \hline
0110 & (2,3) & -1 & (2,3) & -1 & (4) & -1 \\ \hline
0111 & (2,3) & 1 & (2,3) & 1 & (4) & 1 \\ \hline
1000 & (1) & -1 & (1) & -1 & (1,2) & -1 \\ \hline
1001 & (1) & 1 & (1) & 1 & (1,2) & 1 \\ \hline
1010 & (1,3) & -1 & (1,3) & -1 & (3,4) & -1 \\ \hline
1011 & (1,3) & 1 & (1,3) & 1 & (3,4) & 1 \\ \hline
1100 & (1,2) & -1 & (1,2) & -1 & (1,3) & -1 \\ \hline
1101 & (1,2) & 1 & (1,2) & 1 & (1,3) & 1 \\ \hline
1110 & (1,2,3) & -1 & (1,2,3) & -1 & (2,4) & -1 \\ \hline
1111 & (1,2,3) & 1 & (1,2,3) & 1 & (2,4) & 1 \\ \hline
\end{tabular}
$^*$Although we use a table here to describe the three mapping schemes, it is worth noting that the proposed mapping schemes of DTAA-R and DTAA-D can be readily extended to large-scale antenna arrays, which is challenging for LUT with the same design as \cite{osman2015variable, younis2014spatial}.
\end{threeparttable}\\
	\dotfill
\end{table*}

By contrast, when $M$-QAM is employed instead of $M$-PSK, the phase rotation of (\ref{eq2-2-2}) no longer maximises the distance in the constellation. Instead, we propose the following transformation:
\begin{equation}\label{eq2-2-3}
    {T_{QAM}}\left( s \right) = {e^{j\frac{\pi }{{{M_n}}}}} \cdot s,
\end{equation}
where ${M_n} = \max \left\{ {count\left( {\left| {{s_i}} \right| = \left| \chi  \right|} \right)} \right\}$ for $i \in [1,M]$, while $count\left(  \cdot  \right)$ represents the number of occurrences of the event in parentheses, with $\left| \chi  \right|$ representing the set of magnitudes of all QAM constellation points. For example, when the normalised 16QAM constellation is adopted, we have $\left | \chi  \right |=\left \{ \frac{1}{\sqrt{5}},1,\frac{3}{\sqrt{5}} \right \}$, yielding $a_{1}=count\left ( \left | s_{i} \right |=\frac{1}{\sqrt{5}} \right )=4$, $a_{2}=count\left ( \left | s_{i} \right |=1 \right )=8$ and $a_{3}=count\left ( \left | s_{i} \right |=\frac{3}{\sqrt{5}} \right )=4$. Therefore, ${M_n} = 8$ is obtained according to (\ref{eq2-2-3}). Following the same philosophy, we have ${M_n} = 12$ for 64QAM.

We note that although the special case of the all-zero spatial bitstream is used for activating all TAs, other legitimate TA activation patterns that contain at least one activated TA can also be readily chosen for the special case in (\ref{eq2-2-1}), provided that the modulated symbol for the chosen pattern is rotated according to (\ref{eq2-2-2}) and (\ref{eq2-2-3}).

Again, the DTAA-R scheme is exemplified in Fig. \ref{f4} (a), where we have ${N_t} = 3$ TAs and BPSK modulation. More explicitly, let us revisit Fig. \ref{f4} (a) a bit more formally, where the first $m_s = 3$ bits are mapped to the TA activation pattern, while the remaining $m_a = 1$ bit modulates a BPSK symbol. The full set of TA activation patterns is summarised in Table \ref{tab2}. For example, the bit sequence of $\mathbf{b}^{T}=[  1\quad 0 \quad 1 \quad| 0  ]$ results in the TA activation pattern of $\left ( 1,3 \right )$ and the BPSK symbol of $-1$, as seen in Table \ref{tab2}, which results in the transmitted signal vector of $\mathbf{x}=[-1\quad 0\quad -1]^{T}$. Moreover, when the special case of the all-zero spatial bitstream is encounted, the bit sequence of $\mathbf{b}^{T}=[0\quad 0\quad 0\quad|1]$ is mapped to the TA activation pattern of $\left ( 1,2,3 \right )$ and to the rotated BPSK symbol $-i$, as shown in Table \ref{tab2}, which leads to the transmitted signal vector of $\mathbf{x}=[-i\quad -i\quad -i]^{T}$ in Fig. \ref{f4} (a).

\textbf{2) DTAA-D:}
Instead of applying a phase rotation to the APSK constellation, in this section, we propose the DTAA-D mapping method, which assigns a dedicated TA to be uniquely turned on only for the all-zero spatial bitstream. As a result, the on/off states of the $\left ( N_{t}-1 \right )$ TAs are decided by the $\left ( N_{t}-1 \right )$ non-all-zero spatial bits of Fig. \ref{f4} (b) and of Table \ref{tab2} without any ambiguity, hence eliminating the dependence of the DTAA-R scheme on the constellation rotation of (\ref{eq2-2-2}) and (\ref{eq2-2-3}). More explicitly, the DTAA-D mapping rule is defined by
\begin{equation}\label{eq2-2-4}
\left\{ {\begin{array}{*{20}{l}}
{{{\bf{b}}_s^{T}} \ne {\bf{0}} \Leftrightarrow {\bf{x}} = {{\left[ {\begin{array}{*{20}{c}}
{{\bf{b}}_s^{T}}&0
\end{array}} \right]}^T} \times s}\\
{{{\bf{b}}_s^{T}} = {\bf{0}} \Leftrightarrow {\bf{x}} = {{\left[ {\begin{array}{*{20}{c}}
{{{\bf{0}}_{\left( {{N_t} - 1} \right) \times 1}^{T}}}&1
\end{array}} \right]}^T} \times s}.
\end{array}} \right.
\end{equation}
Therefore, the last TA of Fig. \ref{f4} (b) is activated for the special case of ${\bf{b}}_s^{T} = {\bf{0}}$.

Again, the DTAA-D arrangement is exemplified in Fig. \ref{f4} (b), where the incoming data bits are also divided into two parts: a spatial bit sequence ${{\bf{b}}_s^{T}}$ and a data bit sequence ${{\bf{b}}_a^{T}}$. We note that ${N_t} = 4$, $m_s = 3$, and $m_a = 1$ are assumed in Fig. \ref{f4} (b), and the corresponding TA activation patterns are also summarised in Table \ref{tab2}. Specifically, based on Fig. \ref{f4} (b) and Table \ref{tab2}, the information bit sequence of ${\bf{b}}^T = [ 1\quad0\quad1\quad|0 ]$ is transmitted as ${\bf{x}} = {[ { - 1}\quad0\quad{ - 1}\quad0]^T}$. However, based on (\ref{eq2-2-4}), the bit sequence of ${\bf{b}}^T = [ 0\quad0\quad0\quad|1]$ is mapped to $\mathbf{x}=[0\quad 0\quad 0\quad1]^{T}$, as seen in Table \ref{tab2}, which avoids the constellation rotation of $-i$ seen in the DTAA-R example. We will demonstrate in Section \ref{s5} that as the benefit of eliminating the DTAA-R scheme's constellation rotation, the pairwise error probability is increased by the DTAA-D scheme. Nonetheless, we note that this arrangement does not compromise the LCIT-GSM throughput, since a total of $\left ( N_{t}-1 \right )$ spatial bits are transmitted by $N_{t}$ TAs, which is in line with (\ref{eq2-1-2}).

\textbf{3) LUT\footnote{We note that LUT uses the same mapping design as the VGSM of \cite{osman2015variable, younis2014spatial}, and we include it here in order to share the low complexity detectors proposed in Section \ref{s3}.}:}
The DTAA-D arrangement is capable of achieving the full LCIT-GSM throughput without any constellation rotation, but unfortunately, the distance between the legitimate TA activation patterns has not been maximised. More explicitly, according to (\ref{eq2-1-1}), there are a total of $\left ( 2^{N_{t}}-1 \right )$ legitimate TA activation patterns, where the case of deactivating all TAs is eliminated. However, it can be seen in (\ref{eq2-1-2}) that only $2^{N_{t}-1}$ TA activation patterns are needed. Against this background, the LUT arrangement is introduced to further optimize the DTAA-D scheme's spatial constellation, but suffer from the high complexity of the mapping design \cite{osman2015variable, younis2014spatial}.

Specifically, for the case of $N_{t}=4$, there are a total of $2^{N_{t}}-1=15$ possible TA activation patterns, which is constituted by $C_{4}^{1}=4$, $C_{4}^{2}=6$, $C_{4}^{3}=4$ and $C_{4}^{4}=1$ according to (\ref{eq2-1-1}). Nonetheless, only $2^{N_{t}-1}=8$ TA activation patterns are needed. Therefore, the first rule of LUT is to prioritize the TA activation patterns with smaller number $N_{a}$ of activated TAs. For example, the TA activation patterns that only activate a single TA out of $N_{t}=4$ result in the transmitted signal vectors of $\mathbf{x}^{T}=\left \{ \left [ s,0,0,0 \right ],\left [ 0,s,0,0 \right ],\left [ 0,0,s,0 \right ],\left [ 0,0,0,s \right ] \right \}$, where the TA activation pattern always differs in the maximum possible number of 4 positions. Following this, for the TA activation patterns associated with the same $N_{a}$, the pairs that have the highest possible number of different activated TA positions are chosen. For example, after admitting the $C_{4}^{1}=4$ combinations associated with a single activated TA, we still need $2^{N_{t}-1}-C_{4}^{1}=4$ patterns out of the $C_{4}^{2}=6$ combinations associated with $N_{a}=2$ activated TAs. Hence, the TA activation pattern pairs of $\left ( 1,2 \right )$ and $\left ( 3,4 \right )$ are chosen, which always differ in the maximum possible number of 4 positions. Similarly, the pair of $\left ( 1,3 \right )$ and $\left ( 2,4 \right )$ are subsequently chosen. Although $\left ( 1,2 \right )$ and $\left ( 1,3 \right )$ now only differ in 3 activated TA positions, we have ensured that the maximally separated pairs are used.

In summary, the LUT arrangement is exemplified in Fig. \ref{f4} (c) for the case of $N_{t}=4$, $m_{t}=3$ and $m_{a}=1$, where the list of TA activation patterns is also presented in Table \ref{tab2}. The LUT mapping may be represented by
\begin{equation}\label{eq2-2-5}
    {{\bf{b}}_s^{T}} \Leftrightarrow {\bf{x}} = \mathrm{U}\left( {{{\bf{b}}_s}} \right) \times s,
\end{equation}
where $\mathrm{U}\left(  \cdot  \right)$ returns a column vector having entries of $0$ or $1$, and the above LUT mapping rules ensure that the binary vectors $\left \{ \mathrm{U}\left ( \mathbf{b}_{s} \right ) \right \}_{\forall \mathbf{b}_{s}}$ have the maximum possible Hamming distance.

\section{Signal Detection}\label{s3}
In this section, we proceed to devise low-complexity signal detectors for the LCIT-GSM receiver. We first consider the optimal MLD in Section \ref{3_1}, followed by a pair of reduced-complexity detectors in Sections \ref{3_2} and \ref{3_3}, respectively.
\subsection{Maximum Likelihood Detection}\label{3_1}
From the detector's perspective, let us represent the tentative transmitted signal vector as ${{\bf{x}}_{k, l}}$, where the subscript $k$ denotes the $k$-th TA activation pattern\footnote{Note that the subscript $k$ represents the $k$-th TA activation pattern, not the $k$-th TA in SM. The number of active TAs is variable and determined by the $k$-th TA activation pattern.}, while the subscript $l$ denotes the $l$-th $M$-PSK/QAM symbol. We assume that perfect channel knowledge is available for the receiver. As a result, the MLD finds the estimate of $\bf{x}$ by performing a full search over all legitimate indices $k$ and $l$ formulated as \cite{jeganathan2008spatial}:
\begin{equation}\label{eq3-1-1}
    \begin{split}
\left[ {\hat k_{MLD}, \hat l_{MLD}} \right] &= \mathop {\arg \max }\limits_{k,l} {p_{\bf{y}}}\left( {\left. {\bf{y}} \right|{{\bf{x}}_{k,l}},{\bf{H}}} \right)\\
 &= \mathop {\arg \min }\limits_{k,l} {\left\| {{\bf{y}} - {\bf{H}}{{\bf{x}}_{k,l}}} \right\|^2}\\
 &= \mathop {\arg \min }\limits_{k,l} \left \| \mathbf{y}-\mathbf{g}_{k}s_{l} \right \|^{2},
\end{split}
\end{equation}
where the probability density function (PDF) of ${\bf{y}}$ conditioned on ${{{\bf{x}}_{k,l}}}$ and ${\bf{H}}$ is given by:
\begin{equation}\label{eq3-1-2}
{p_{\bf{y}}}\left( {\left. {\bf{y}} \right|{{\bf{x}}_{k,l}},{\bf{H}}} \right) = \frac{1}{{{{\left( {\pi \sigma _n^2} \right)}^{{N_r}}}}}\exp \left( { - \frac{{{{\left\| {{\bf{y}} - {\bf{H}}{{\bf{x}}_{k,l}}} \right\|}^2}}}{{\sigma _n^2}}} \right),
\end{equation}
${s_l}$ represents the $l$-th symbol of the constellation, and $\mathbf{g}_{k}$ is the $k$-th column of the equivalent $N \times {N_r}$ LCIT-GSM channel matrix $\mathbf{G}$ from the perspective of the APSK symbol, where we have $N = {2^{{N_t}}}$ for DTAA-R and $N = {2^{{N_t} - 1}}$ for DTAA-D and LUT. In summary, the equivalent channel matrix $\mathbf{G}$ is defined as follows
\begin{equation}\label{eq3-1-3}
\left\{ {\begin{array}{*{20}{l}}
\mathbf{G}_{DTAA-R}=\begin{bmatrix}
\mathbf{H}\mathbf{b}_{s,1} & \cdots & \mathbf{H}\mathbf{b}_{s,N-1}  & \mathbf{H}\mathbf{1}e^{j\frac{\pi }{M_{n}}}
\end{bmatrix}\\
\mathbf{G} _{DTAA-D}=\begin{bmatrix}
\mathbf{H}\begin{bmatrix}
\mathbf{b}_{s,1}\\ 0

\end{bmatrix} & \cdots  & \mathbf{H}\begin{bmatrix}
\mathbf{b}_{s,N-1}\\ 0

\end{bmatrix} & \mathbf{H}\begin{bmatrix}
\mathbf{0}\\ 1

\end{bmatrix}
\end{bmatrix}\\
\mathbf{G}_{LUT}=\begin{bmatrix}
\mathbf{H}\mathrm{U}\left ( \mathbf{b}_{s,0} \right ) & \mathbf{H}\mathrm{U}\left ( \mathbf{b}_{s,1} \right ) & \cdots  & \mathbf{H}\mathrm{U}\left ( \mathbf{b}_{s,N-1} \right )
\end{bmatrix},
\end{array}} \right.
\end{equation}
where $M_n = M$ for $M$-PSK, and ${\bf{b}}_{s,0} = \bf{0}$ denotes the spatial bitstream with all zeros, while ${\bf{b}}_{s,1}, \cdots, {\bf{b}}_{s,N-1}$ denote the other spatial bitstreams.

As a result, it can be readily seen from (\ref{eq3-1-1}) that the LCIT-GSM system is now equivalent to an SM system with $N$ TAs, where all of the three mapping arrangements of DTAA-R, DTAA-D and LUT have the same detector structure.
\subsection{Two-stage near-Maximum Likelihood Detection}\label{3_2}
The MLD has to jointly search through $N$ TA activation patterns and $M$ symbols, hence its complexity order is given by $\mathcal{O}\left( {MN} \right)$, which grows exponentially both with the number $N_{t}$ of TAs and number $m_{s}$ of modulated bits. {\color{black}{In this section, we conceive the low-complexity two-stage near-MLD (TMLD) scheme, which is similar to the massive MIMO detector of \cite{choi2016near} and to the sphere decoder of \cite{younis2013generalised}. This two-stage TMLD first detects the modulated symbol index $l$ from the reduced search space as follows:}}
\begin{equation}\label{eq3-2-1}
{\hat l} = \mathop {\arg \min }\limits_l \left \| \mathbf{y}-\mathbf{h}_{k'}s_{l} \right \|^{2}.
\end{equation}
We note that the range of $k'$ is given by $k' \in \left\{ {1,2, \cdots ,{N_t}} \right\}$, which is reduced from $k\in \left \{ 1,2,\cdots ,N \right \}$.

Then, the TMLD relaxes the range of $l$ appropriately based on the estimated ${\hat l}$, which means several $l$ values can be selected based on (\ref{eq3-2-1}), instead of a single ${\hat l}$, since a single one would unlikely to be the globally optimal one. Therefore, TMLD defines the reduced candidate set of $l$ using ${\hat l}$ in (\ref{eq3-2-1}) as
\begin{equation}\label{eq3-2-2}
{\chi _{TMLD}} = \left\{ {\left. l \right|\frac{\left \| \mathbf{y}-\mathbf{h}_{k'}s_{l} \right \|^{2}}{\left \| \mathbf{y}-\mathbf{h}_{k'}s_{\hat{l}} \right \|^{2}} \le c} \right\},
\end{equation}
where $c \ge 1$ is a constant that controls the size of the set ${\chi _{TMLD}}$.

Following this, the second stage of TMLD is performed based on (\ref{eq3-1-1}) as:
\begin{equation}\label{eq3-2-3}
    \left[ {\hat k_{TMLD}, \hat l_{TMLD}} \right] = \mathop {\arg \min }\limits_{k,l \in \chi _{TMLD} } \left \| \mathbf{y}-\mathbf{g}_{k}s_{l} \right \|^{2}.
\end{equation}

The MLD guarantees that the solution is optimal by searching through all the elements in the available set. By contrast, TMLD searches for $k$ and $l$ in a reduced region by relaxing the rough estimate of $l$, which is similar to the sphere decoding philosophy of \cite{younis2013generalised, younis2011sphere}. When $c$ is infinite, TMLD becomes equivalent to MLD.
\subsection{Decoupling Based Maximum Likelihood Detection}\label{3_3}
In this subsection, we propose the novel decoupling based MLD (DMLD), which firstly obtains the optimum modulation indices for all TA activation pattern candidates and then detects the optimal TA activation pattern with the aid of the demodulated $M$-APSK symbol. More explicitly, for each tentative TA combination index $k$, the LCIT-GSM detection in (\ref{eq3-1-1}) can be simplified to
\begin{equation}\label{eq3-3-1}
    \begin{split}
{\hat l_{k}}  &= \mathop {\arg \min }\limits_{l} \left\{{\left\|{\mathbf{g }}_{k}\right\|}^{2}{s}_{l}-2\Re\left\{{s}_{l}^{*}{\mathbf{g }}_{k}^{H}\mathbf{y}\right\}\right\}\\
  &= \mathop {\arg \min }\limits_{l} \left\{{\left|{s}_{l}-{p}_{k}\right|}^{2}-{\left|{p}_{k}\right|}^{2}\right\}\\
  &= \mathop {\arg \min }\limits_{l} \left\{{\left|{s}_{l}-{p}_{k}\right|}^{2}\right\},
\end{split}
\end{equation}
where the demodulator's decision variable is given by ${p}_{k}=\frac{{\mathbf{g} }_{k}^{H}\mathbf{y}}{{\left\|{\mathbf{g} }_{k}\right\|}^{2}}$. {\color{black}{In this way, the optimum modulated symbol index $\hat{l}_{k}$ associated with all TA activation indices may be directly obtained by demapping $p_{k}$ to the closest constellation point, assuming that a regular APSK constellation is used.}}

When $M$-PSK is employed, the demodulation of (\ref{eq3-3-1}) may be performed by rounding the phase of the decision variable to the nearest $M$-PSK index as \cite{xu2017two}\footnote{In DMLD, we estimate $s$ instead of its index, since $s$ can be directly obtained by the rounding operation.}
\begin{equation}\label{eq3-3-2}
\hat{s}_{k}= \exp\left \{ j\left [ \frac{\pi }{2}\left \lfloor \left(\angle {p}_{k}-\phi_{0} \right)\frac{2}{\pi } \right \rceil+\phi_{0}  \right ] \right \},
\end{equation}
where $\phi_{0}$ is the phase of the first $M$-PSK symbol associated with all-zero bits.

Similarly, as shown in \cite{xu2017two, rajashekar2013reduced}, the QAM constellation may be decoupled into two real-valued APSK signal sets as $s={s}_{I}+j{s}_{Q}$. Since ${s}_{I}$ and ${s}_{Q}$ are from orthogonal dimensions, we have
\begin{equation}\label{eq3-3-3}
\hat{l}_{k}=\underset{l}{\arg \min}\left [ \left ( s_{I}-\Re \left ( p_{k} \right ) \right )^{2}+\left ( s_{Q}-\Im \left ( p_{k} \right ) \right )^{2} \right ],
\end{equation}
which may be further decoupled into
\begin{equation}\label{eq3-3-4}
    \Re \left ( \hat{s}_{k} \right )= \min\left [ \max\left ( 2\left \lfloor \frac{\alpha \Re \left ( p_{k} \right )+1}{2} \right \rceil-1,-M+1 \right ),M-1 \right ],
\end{equation}
and
\begin{equation}\label{eq3-3-5}
    \Im \left ( \hat{s}_{k} \right )= \min\left [ \max\left ( 2\left \lfloor \frac{\alpha \Im \left ( p_{k} \right )+1}{2} \right \rceil-1,-M+1 \right ),M-1 \right ].
\end{equation}
We note that $\alpha =\sqrt{\frac{\sum_{m=0}^{\frac{\sqrt{M}}{2}-1}\left ( \sqrt{M}-2m-1 \right )^{2}}{\frac{\sqrt{M}}{4}}}$ is the normalization factor of the QAM constellation that yields $\hat{s}_{k}=\frac{1}{\alpha }\left [ \Re \left ( \hat{s}_{k} \right )+\Im \left ( \hat{s}_{k} \right ) \right ]$ and ensures unity average symbol power \cite{xu2017two, rajashekar2013reduced}.

From (\ref{eq3-3-2}), (\ref{eq3-3-4}) and (\ref{eq3-3-5}) we obtain the total number $N$ of demodulated symbols $S_{min}=\left \{ \hat{s}_{1},\hat{s}_{2},\cdots ,\hat{s}_{N} \right \}$ associated with the $N$ hypotheses, for all $1\leq k\leq N$. Thus, we have
\begin{equation}\label{eq3-3-6}
\begin{split}
    \left [ \hat{k}_{DMLD}, \hat{s}_{DMLD} \right ] &=\underset{k,s}{\arg \min}\left \| \mathbf{y}-\mathbf{g}_{k}s_{l} \right \|^{2}\\
    &=\underset{k}{\arg \min}\left \| \mathbf{y}-\mathbf{g}_{k}\hat{s}_{k} \right \|^{2}.
\end{split}
\end{equation}
{\color{black}{Once ${\hat k_{DMLD}}$ is estimated, we have ${\hat s_{DMLD}} = {\hat s_{{{\hat k}_{DMLD}}}}$, which is the ${\hat k_{DMLD}}$-th element of $S_{min}$.}}

The DMLD's complexity order does not grow with the number $M$ of modulation levels, which constitutes the most appealing advantages of the optimal LCIT-GSM detection, especially when employing high-order $M$-PSK/QAM schemes. More importantly, the DMLD of (\ref{eq3-3-1})-(\ref{eq3-3-6}) substantially reduces the detection complexity without imposing any performance loss on the MLD of (\ref{eq3-1-1}).

\section{Performance Analysis}\label{s4}
{\color{black}{In this section, we present analytical results for characterising the error performance of the uncoded LCIT-GSM system, where the MLD is assumed to obtain the optimal reference bound\footnote{\color{black}{We note that the theoretical performance bounds are different for different detectors. Here we derive the ultimate theoretical bound for the MLD.}}.}} A tight UB on the BEP is derived. Additionally, we also analyze the computational complexity of the proposed LCIT-GSM detectors.
\subsection{Uncoded Performance}\label{s4-1}
\subsubsection{Classic Union Bound}
First of all, the analytical performance of our LCIT-GSM system is evaluated using the well-known union bounding technique. The average BEP of LCIT-GSM is given by:
\begin{equation}\label{eq4-1-1}
P_{b}\leq \frac{1}{MN}\sum_{k=1}^{N}\sum_{l=1}^{M}\sum_{\tilde{k}=1}^{N}\sum_{\tilde{l}=1}^{M}N\left ( \mathbf{x}_{k,l} \rightarrow \mathbf{x}_{\tilde{k},\tilde{l}}  \right )P\left ( \mathbf{x}_{k,l} \rightarrow \mathbf{x}_{\tilde{k},\tilde{l}}  \right ),
\end{equation}
where $\mathbf{x}_{k,l}$ indicates the symbol $s_l$ transmitted using the $k$-th TA activation pattern, while $N\left ( \mathbf{x}_{k,l} \rightarrow \mathbf{x}_{\tilde{k},\tilde{l}}  \right )$ is the number of bits in error between $\mathbf{x}_{k,l}$ and $\mathbf{x}_{\tilde{k},\tilde{l}}$, and $P\left ( \mathbf{x}_{k,l}\rightarrow \mathbf{x}_{\tilde{k},\tilde{l}}  \right )$ denotes the probability of deciding on $\mathbf{x}_{\tilde{k},\tilde{l}}$, given that $\mathbf{x}_{k,l}$ is transmitted.

Following this, the probability $P\left ( \mathbf{x}_{k,l}\rightarrow \mathbf{x}_{\tilde{k},\tilde{l}}  \right )$ can be computed by using (\ref{eq3-1-2}) as follows
\begin{equation}\label{eq4-1-2}
\begin{split}
    P\left ( \mathbf{x}_{k,l}\rightarrow \mathbf{x}_{\tilde{k},\tilde{l}}  \right )&=P\left ( \left \| \mathbf{y}-{\bf{g}}_{k}s_{l} \right \|^{2}> \left \| \mathbf{y}-{\bf{g}} _{\tilde{k}}s_{\tilde{l}} \right \|^{2} \right )\\
    &=Q\left ( \sqrt{\frac{1}{2\sigma _{n}^{2}}\left \| {\bf{g}}_{k}s_{l}-{\bf{g}} _{\tilde{k}}s_{\tilde{l}} \right \|^{2}} \right ).
\end{split}
\end{equation}

According to Eq. (14-4-15) in \cite{proakis2001digital}, we obtain
\begin{align}\label{eq4-1-3}
{P_b} &\le \frac{1}{{MN}}\sum\limits_{k = 1}^N {\sum\limits_{l = 1}^M {\sum\limits_{\tilde k = 1}^N {\sum\limits_{\tilde l = 1}^M {} } } } \notag\\
 & \quad \times N\left( {{{\bf{x}}_{k,l}} \to {{\bf{x}}_{\tilde k,\tilde l}}} \right)\mu _\alpha ^{{N_r}}\sum\limits_{n = 0}^{{N_r} - 1} {C_{{N_r} - 1 + n}^n{{\left( {1 - {\mu _\alpha }} \right)}^n}} ,
\end{align}
with
\begin{equation}\label{eq4-1-4}
    {\mu _\alpha } =\mu \left ( \sigma _{\alpha }^{2} \right )= \frac{1}{2}\left( {1 - \sqrt {\frac{{\sigma _\alpha ^2}}{{1 + \sigma _\alpha ^2}}} } \right),
\end{equation}
and
\begin{equation}\label{eq4-1-5}
    \sigma _\alpha ^2 = \frac{1 }{4\sigma _{n}^{2}}\left( {{{\left| {{s_l}} \right|}^2} + {{\left| {{s_{\tilde l}}} \right|}^2}} \right).
\end{equation}
\subsubsection{Improved Upper Bound}
The theoretical UB above is derived based on the conventional SM study of \cite{proakis2001digital}, which does not take into account the specific effects of the spatial constellation and symbol constellation. To avoid these limitations, an improved UB is proposed in \cite{di2012bit}. In this subsection, we will derive our improved performance bound for LCIT-GSM based on the methodology of \cite{di2012bit}. We firstly offer \emph{Proposition 1} that is conceived for the sake of improving the UB of our LCIT-GSM system, and then we further derive a simplified UB under Rayleigh flat fading from \emph{Proposition 1}.

\emph{Proposition 1:} The average BEP of LCIT-GSM can be tightly upper bounded as follows:
\begin{equation}\label{eq4-1-6}
{P_b} \le {P_{b,signal}} + {P_{b,spatial}} + {P_{b,joint}},
\end{equation}
where ${P_{b,signal}}$, ${P_{b,spatial}}$, ${P_{b,joint}}$ are defined as
\begin{equation}\label{eq4-1-7}
\left\{ {\begin{array}{*{20}{l}}
{{P_{b,signal}} = \frac{{{{\log }}M}}{{{N}{{\log }}\left( {M{N}} \right)}}\sum\limits_{k = 1}^{{N}} {{P_{b, MOD}}\left( k \right)} }\\
{{P_{b,spatial}} = \frac{{{{\log }}{N}}}{{M{{\log }}\left( {M{N}} \right)}}\sum\limits_{l = 1}^M {{P_{b,SSK}}\left( l \right)} }\\
{{P_{b,joint}} = \frac{{\sum\limits_{k = 1}^{{N}} {\sum\limits_{l = 1}^M {\sum\limits_{\tilde k = 1}^{{N}} {\sum\limits_{\tilde l = 1}^M {N\left( {\mathbf{x}_{k,l} \to {\mathbf{x}_{\tilde{k},\tilde{l}}},{s_{\tilde l}}} \right)\Upsilon \left( {\mathbf{x}_{k,l},\mathbf{x}_{\tilde{k},\tilde{l}}} \right)} } } } }}{{M{N}{{\log }}\left( {M{N}} \right)}}},
\end{array}} \right.
\end{equation}
with
\begin{equation}\label{eq4-1-8}
\left\{ {\begin{array}{*{20}{l}}
{{P_{b, MOD}}\left( k \right) = \frac{{\sum\limits_{l = 1}^M {\sum\limits_{\tilde l = 1}^M {\left[ {N\left( {{s_l} \to {s_{\tilde l}}} \right){E_{{\mathbf{g} _k}}}\left\{ P\left ( s_{l}=s_{\tilde{l}} \right ) \right\}} \right]} } }}{{M{{\log }}\left( M \right)}}}\\
{{P_{b,SSK}}\left( l \right) = \frac{{\sum\limits_{k = 1}^{{N}} {\sum\limits_{\tilde k = 1}^{{N}} {N\left( {\mathbf{b}_{s,k} \to {\mathbf{b}_{s,\tilde{k}}}} \right){\Psi _l}\left( {\mathbf{b}_{s,k},\mathbf{b}_{s,\tilde{k}}} \right)} } }}{{{N}{{\log }}\left( {{N}} \right)}}},
\end{array}} \right.
\end{equation}
where
\begin{equation}\label{eq4-1-9}
\left\{ {\begin{array}{*{20}{l}}
{{\Psi _l}\left( {\mathbf{b}_{s,k},\mathbf{b}_{s,\tilde{k}}} \right) = \frac{1}{\pi }\int\limits_0^{{\pi  \mathord{\left/
 {\vphantom {\pi  2}} \right.
 \kern-\nulldelimiterspace} 2}} {{{\mathcal{M}}_{\gamma \left( {\mathbf{b}_{s,k},\mathbf{b}_{s,\tilde{k}}} \right)}}\left( {\frac{{ {\left| {{s_l}} \right|^2}}}{{4\sigma _{n}^{2}{{\sin }^2}\theta }}} \right)d\theta } }\\
{\Upsilon \left( {\mathbf{x}_{k,l},\mathbf{x}_{\tilde{k},\tilde{l}}} \right) = \frac{1}{\pi }\int\limits_0^{{\pi  \mathord{\left/
 {\vphantom {\pi  2}} \right.
 \kern-\nulldelimiterspace} 2}} {{{\mathcal{M}}_{\gamma \left( {\mathbf{x}_{k,l},\mathbf{x}_{\tilde{k},\tilde{l}}} \right)}}\left( {\frac{1 }{{4\sigma _{n}^{2}{{\sin }^2}\theta }}} \right)d\theta } },
\end{array}} \right.
\end{equation}
and
\begin{equation}\label{eq4-1-10}
\left\{ {\begin{array}{*{20}{l}}
\gamma \left( {\mathbf{b}_{s,k},{\mathbf{b}_{s,\tilde{k}}}} \right) = \left \| \mathbf{g}_{k}-\mathbf{g}_{\tilde{k}} \right \|^{2}\\
\gamma \left( {\mathbf{x}_{k,l},\mathbf{x}_{\tilde{k},\tilde{l}}} \right) = \left \| \mathbf{g}_{k}s_{l}-\mathbf{g}_{\tilde{k}}s_{\tilde{l}} \right \|^{2}.
\end{array}} \right.
 \end{equation}

\emph{Proof:} Please refer to Appendix A in \cite{di2012bit}.

{\color{black}{As shown in (\ref{eq4-1-9}), in order to obtain the ABEP upper bound described in (\ref{eq4-1-6}), we have to derive the MGFs of $\gamma \left( {\mathbf{b}_{s,k},{\mathbf{b}_{s,\tilde{k}}}} \right) $ and $\gamma \left( {\mathbf{x}_{k,l},\mathbf{x}_{\tilde{k},\tilde{l}}} \right)$. According to \cite{younis2014spatial, liu2019spatial}, we know that
\begin{equation}\label{eq4-1-11}
    \begin{split}
\gamma \left( {{{\bf{b}}_{s,k}},{{\bf{b}}_{s,\tilde k}}} \right) &= {\left\| {{{\bf{g}}_k} - {{\bf{g}}_{\tilde k}}} \right\|^2} = {\left\| {{\bf{G}}\left( {{{\bf{e}}_k} - {{\bf{e}}_{\tilde k}}} \right)} \right\|^2}\\
 &= \sum\limits_{n = 1}^{{N_r}} {{{\bf{G}}_{n,:}}\left( {{{\bf{e}}_k} - {{\bf{e}}_{\tilde k}}} \right){{\left( {{{\bf{e}}_k} - {{\bf{e}}_{\tilde k}}} \right)}^H}{\bf{G}}_{n,:}^H} \\
 &= {{{\bf{\mathord{\buildrel{\lower3pt\hbox{$\scriptscriptstyle\smile$}} 
\over g} }}}^T}\left[ {{{\bf{I}}_N} \otimes \left( {{{\bf{e}}_k} - {{\bf{e}}_{\tilde k}}} \right){{\left( {{{\bf{e}}_k} - {{\bf{e}}_{\tilde k}}} \right)}^H}} \right]{{{\bf{\mathord{\buildrel{\lower3pt\hbox{$\scriptscriptstyle\smile$}} 
\over g} }}}^ * },
\end{split}
\end{equation}
and
\begin{equation}\label{eq4-1-12}
    \begin{split}
\gamma \left( {{{\bf{x}}_{k,l}},{{\bf{x}}_{\tilde k,\tilde l}}} \right) &= {\left\| {{{\bf{g}}_k}{s_l} - {{\bf{g}}_{\tilde k}}{s_{\tilde l}}} \right\|^2} = {\left\| {{\bf{G}}\left( {{{\bf{x}}_{k,l}} - {{\bf{x}}_{\tilde k,\tilde l}}} \right)} \right\|^2}\\
 &= \sum\limits_{n = 1}^{{N_r}} {{{\bf{G}}_{n,:}}\left( {{{\bf{x}}_{k,l}} - {{\bf{x}}_{\tilde k,\tilde l}}} \right){{\left( {{{\bf{x}}_{k,l}} - {{\bf{x}}_{\tilde k,\tilde l}}} \right)}^H}{\bf{G}}_{n,:}^H} \\
 &= {{{\bf{\mathord{\buildrel{\lower3pt\hbox{$\scriptscriptstyle\smile$}} 
\over g} }}}^T}\left[ {{{\bf{I}}_N} \otimes \left( {{{\bf{x}}_{k,l}} - {{\bf{x}}_{\tilde k,\tilde l}}} \right){{\left( {{{\bf{x}}_{k,l}} - {{\bf{x}}_{\tilde k,\tilde l}}} \right)}^H}} \right]{{{\bf{\mathord{\buildrel{\lower3pt\hbox{$\scriptscriptstyle\smile$}} 
\over g} }}}^ * },
\end{split}
\end{equation}
where ${{{\bf{G}}_{n,:}}}$ represents the $n$-th row of ${\bf{G}}$, ${{\bf{e}}_k}$ is the ${N_t} \times 1$ unit vector with only the $k$-th entry is $1$, and ${\bf{\mathord{\buildrel{\lower3pt\hbox{$\scriptscriptstyle\smile$}} 
\over g} }} = vec\left( {{{\bf{G}}^T}} \right)$ is a ${N_r}N \times 1$ vector obtained by the vectorization of ${{\bf{G}}^T}$. In this paper, we explicitly investigate the canonical i.i.d. Rayleigh fading scenarios. Nevertheless, the equivalent LCIT-GSM channels are correlated. Hence, the MGFs of $\gamma \left( {\mathbf{b}_{s,k},{\mathbf{b}_{s,\tilde{k}}}} \right) $ and $\gamma \left( {\mathbf{x}_{k,l},\mathbf{x}_{\tilde{k},\tilde{l}}} \right)$ can be derived in the same way as in \cite{younis2014spatial, liu2019spatial}, which can be expressed as
\begin{align}\label{eq4-1-13}
{M_{\gamma \left( {{{\bf{b}}_{s,k}},{{\bf{b}}_{s,\tilde k}}} \right)}}\left( t \right) &= \det \left[ {{{\bf{I}}_{{N_r}N}} + t{{\bf{C}}_{{\bf{\mathord{\buildrel{\lower3pt\hbox{$\scriptscriptstyle\smile$}} 
\over g} }}}}\left( {{{\bf{I}}_N}} \right.} \right. \notag\\
&\quad {\left. {\left. { \otimes \left( {\left( {{{\bf{e}}_k} - {{\bf{e}}_{\tilde k}}} \right){{\left( {{{\bf{e}}_k} - {{\bf{e}}_{\tilde k}}} \right)}^H}} \right)} \right)} \right]^{ - 1}},
\end{align}
and
\begin{align}\label{eq4-1-14}
{M_{\gamma \left( {{{\bf{x}}_{k,l}},{{\bf{x}}_{\tilde k,\tilde l}}} \right)}}\left( t \right) &= \det \left[ {{{\bf{I}}_{{N_r}N}} + t{{\bf{C}}_{{\bf{\mathord{\buildrel{\lower3pt\hbox{$\scriptscriptstyle\smile$}} 
\over g} }}}}\left( {{{\bf{I}}_N}} \right.} \right. \notag\\
&\quad {\left. {\left. { \otimes \left( {\left( {{{\bf{x}}_{k,l}} - {{\bf{x}}_{\tilde k,\tilde l}}} \right){{\left( {{{\bf{x}}_{k,l}} - {{\bf{x}}_{\tilde k,\tilde l}}} \right)}^H}} \right)} \right)} \right]^{ - 1}},
\end{align}
where $\det \left( {\bf{S}} \right)$ denotes the determinant of matrix ${\bf{S}}$, ${{\bf{C}}_{{\bf{\mathord{\buildrel{\lower3pt\hbox{$\scriptscriptstyle\smile$}} 
\over g} }}}} = \frac{{\rm{1}}}{{{N_r}}}{{\bf{I}}_{{N_r}}} \otimes E\left( {{{\bf{G}}^T}{{\bf{G}}^ * }} \right)$ is the covariance matrix of ${\bf{\mathord{\buildrel{\lower3pt\hbox{$\scriptscriptstyle\smile$}} 
\over g} }}$, depending on the specific spatial mapping scheme introduced in Section \ref{2_2}. More specifically, the $\left( {k,\tilde k} \right)$ element of ${{\bf{C}}_{{{\bf{G}}^T}}} = \frac{1}{{{N_r}}}E\left( {{{\bf{G}}^T}{{\bf{G}}^ * }} \right)$ is equal to $1$ when the $k$-th and $\tilde k$-th TA activation patterns share more than one activated TA, and $0$ for the rest.

After substituting (\ref{eq4-1-13}) and (\ref{eq4-1-14}) into (\ref{eq4-1-7}) $ \sim $ (\ref{eq4-1-9}), we arrive at
\begin{align}\label{eq4-1-15}
{P_{b,spatial}} &= \frac{1}{{\pi MN{{\log }}\left( {MN} \right)}}\sum\limits_{l = 1}^M {\sum\limits_{k = 1}^N {\sum\limits_{\tilde k = 1}^N {N\left( {{{\bf{b}}_{s,k}} \to {{\bf{b}}_{s,\tilde k}}} \right)} } } \notag\\
 &\times \int\limits_0^{{\pi  \mathord{\left/
 {\vphantom {\pi  2}} \right.
 \kern-\nulldelimiterspace} 2}} {\det \left[ {{{\bf{I}}_{{N_r}N}} + \frac{{{{\left| {{s_l}} \right|}^2}}}{{4\sigma _n^2{{\sin }^2}\theta }}{{\bf{C}}_{{\bf{\mathord{\buildrel{\lower3pt\hbox{$\scriptscriptstyle\smile$}} 
\over g} }}}}} \right.} \notag\\
&\quad  \times {\left. {\left( {{{\bf{I}}_N} \otimes \left( {\left( {{{\bf{e}}_k} - {{\bf{e}}_{\tilde k}}} \right){{\left( {{{\bf{e}}_k} - {{\bf{e}}_{\tilde k}}} \right)}^H}} \right)} \right)} \right]^{ - 1}}d\theta  \notag\\
 &\le \frac{1}{{2\pi MN{{\log }}\left( {MN} \right)}}\sum\limits_{l = 1}^M {\sum\limits_{k = 1}^N {\sum\limits_{\tilde k = 1}^N {N\left( {{{\bf{b}}_{s,k}} \to {{\bf{b}}_{s,\tilde k}}} \right)} } } \notag\\
 &\times \det \left[ {{{\bf{I}}_{{N_r}N}} + \frac{{{{\left| {{s_l}} \right|}^2}}}{{4\sigma _n^2}}{{\bf{C}}_{{\bf{\mathord{\buildrel{\lower3pt\hbox{$\scriptscriptstyle\smile$}} 
\over g} }}}}} \right. \notag\\
 &\quad  \times {\left. {\left( {{{\bf{I}}_N} \otimes \left( {\left( {{{\bf{e}}_k} - {{\bf{e}}_{\tilde k}}} \right){{\left( {{{\bf{e}}_k} - {{\bf{e}}_{\tilde k}}} \right)}^H}} \right)} \right)} \right]^{ - 1}},
\end{align}
and
\begin{align}\label{eq4-1-16}
{P_{b,joint}} &= \frac{1}{{\pi MN{{\log }}\left( {MN} \right)}}\sum\limits_{k = 1}^N {\sum\limits_{l = 1}^M {\sum\limits_{\tilde k = 1}^N {\sum\limits_{\tilde l = 1}^M {N\left( {{{\bf{x}}_{k,l}} \to {{\bf{x}}_{\tilde k,\tilde l}}} \right)} } } } \notag \\
 &\times \int\limits_0^{{\pi  \mathord{\left/
 {\vphantom {\pi  2}} \right.
 \kern-\nulldelimiterspace} 2}} {\det \left[ {{{\bf{I}}_{{N_r}N}} + \frac{1}{{4\sigma _n^2{{\sin }^2}\theta }}{{\bf{C}}_{{\bf{\mathord{\buildrel{\lower3pt\hbox{$\scriptscriptstyle\smile$}} 
\over g} }}}}} \right.} \notag \\
 & \quad  \times {\left. {\left( {{{\bf{I}}_N} \otimes \left( {\left( {{{\bf{x}}_{k,l}} - {{\bf{x}}_{\tilde k,\tilde l}}} \right){{\left( {{{\bf{x}}_{k,l}} - {{\bf{x}}_{\tilde k,\tilde l}}} \right)}^H}} \right)} \right)} \right]^{ - 1}}d\theta  \notag \\
 &\le \frac{1}{{2\pi MN{{\log }}\left( {MN} \right)}}\sum\limits_{k = 1}^N {\sum\limits_{l = 1}^M {\sum\limits_{\tilde k = 1}^N {\sum\limits_{\tilde l = 1}^M {N\left( {{{\bf{x}}_{k,l}} \to {{\bf{x}}_{\tilde k,\tilde l}}} \right)} } } } \notag \\
 &\times \det \left[ {{{\bf{I}}_{{N_r}N}} + \frac{1}{{4\sigma _n^2}}{{\bf{C}}_{{\bf{\mathord{\buildrel{\lower3pt\hbox{$\scriptscriptstyle\smile$}} 
\over g} }}}}} \right. \notag\\
 & \quad  \times {\left. {\left( {{{\bf{I}}_N} \otimes \left( {\left( {{{\bf{x}}_{k,l}} - {{\bf{x}}_{\tilde k,\tilde l}}} \right){{\left( {{{\bf{x}}_{k,l}} - {{\bf{x}}_{\tilde k,\tilde l}}} \right)}^H}} \right)} \right)} \right]^{ - 1}},
\end{align}
where the Chernoff bound is used for simplifying the numerical integrals \cite{younis2014spatial}.}}

On the other hand, ${P_{b,signal}}$ can also be further simplified to
\begin{equation}\label{eq4-1-17}
    {P_{b,signal}} = \frac{{{{\log }}M}}{{{{\log }}\left( {MN} \right)}}P_{b, MOD}^{Rayleigh}.
\end{equation}

\begin{table*}[!t]
\renewcommand\arraystretch{1.5}
    	\centering
	\caption{\label{tab3} \color{black}{Contrast between three detectors for LCIT-GSM system.}}
\begin{tabular}{|c|c|c|c|c|c|}
\hline
\textit{} & \textbf{BER performance} & \textbf{Order of complexity} & \textbf{Computational complexity} \\ \hline
\textbf{MLD} & optimal & ${\mathcal{O}}\left( {MN} \right)$ & ${\delta_{MLD}} = 6M{N_r}N$ \\ \hline
\textbf{TMLD} & sub-optimal & ${\mathcal {O}}\left( {N} \right)$, for a small $c$ and $M$ & ${\delta_{TMLD}} = 6M{N_r}{N_t} + 6\beta M{N_r}N$ \\ \hline
\textbf{DMLD} & optimal & ${\mathcal {O}}\left( {N} \right)$ & $    {\delta _{DMLD}} = \left\{ {\begin{array}{*{20}{l}}
{\left( {6{N_r} + 10} \right)N,}&{PSK}\\
{\left( {6{N_r} + 12} \right)N,}&{QAM}
\end{array}} \right.$\\ \hline
\end{tabular}
\end{table*}
For PSK and QAM constellations, $P_{b, MOD}^{Rayleigh}$ can be expressed similarly to \cite{simon2005digital} as
\begin{equation}\label{eq4-1-18}
P_{b, PSK}^{Rayleigh} \simeq \frac{{2\sum\limits_{k = 1}^{\max \left( {{M \mathord{\left/
 {\vphantom {M 4}} \right.
 \kern-\nulldelimiterspace} 4},1} \right)} {R\left( \frac{{ {{\sin }^2} {\frac{{\left( {2k - 1} \right)\pi }}{M}} }}{2\sigma _{n}^{2}} \right)} }}{{\max \left( {{{\log }}M,2} \right)}},
\end{equation}
and
\begin{equation}\label{eq4-1-19}
P_{b, QAM}^{Rayleigh} = \frac{4}{{\sqrt M {{\log }}M}}\sum\limits_{l = 1}^{\frac{1}{2}{{\log }}M} {\sum\limits_{k = 0}^{\left( {1 - {2^{ - l}}} \right)\sqrt M  - 1} {{S_{l,k}}} },
\end{equation}
where 
\begin{equation}\label{eq4-1-20}
    {S_{l,k}} = {\left( { - 1} \right)^{\left\lfloor {\frac{{{2^{l - 1}}k}}{{\sqrt M }}} \right\rfloor }}\left( {{2^{l - 1}} - \left\lfloor {\frac{{{2^{l - 1}}k}}{{\sqrt M }} + \frac{1}{2}} \right\rfloor } \right)R\left( {\frac{{3 {{\left( {2k + 1} \right)}^2}}}{4\sigma _{n}^{2}\left ( M-1 \right )}} \right).
\end{equation}
and
\begin{equation}\label{eq4-1-21}
R\left( x \right) = {\left( {\mu \left( x \right)} \right)^{{N_r}}}\sum\limits_{{n_r} = 0}^{{N_r} - 1} {\left\{ {C_{{N_r} - 1 + {n_r}}^{{n_r}}{{\left( {1 - \mu \left( x \right)} \right)}^{{n_r}}}} \right\}},
\end{equation}

The formulas in (\ref{eq4-1-15}), (\ref{eq4-1-16}) and (\ref{eq4-1-17}) provide useful insights into the performance of LCIT-GSM systems. For example, (\ref{eq4-1-16}) shows that the bit mapping of the PSK/QAM constellation diagram plays an important role in ${P_{b,joint}}$. In particular, while conventional bit mappings based on the Euclidean distance of the signal constellations turn out to be optimal for minimising ${P_{b,signal}}$, additional constraints may be imposed onto the best choice of the signal constellation diagram and onto the related bit mapping for minimising ${P_{b,signal}}$ and ${P_{b,joint}}$ simultaneously.

\subsection{\color{black}{Complexity Analysis}}\label{s4-2}
{\color{black}{In this part, we compare the computational complexity of the detectors proposed for our LCIT-GSM scheme. The detection complexity is quantified in terms of the number of real-valued multiplications required. The complexity of GSM in \cite{younis2010generalised} is given by ${\delta_{GSM}} = 6M{N_r}{2^{\left\lfloor {{{\log }}C_{{N_t}}^{{N_a}}} \right\rfloor }}$. Similarly, for the proposed LCIT-GSM scheme, the detection complexity of the full-search based MLD of (\ref{eq3-1-1}) is given by
\begin{equation}\label{eq4-2-2}
    {\delta_{MLD}} = 6M{N_r}N.
\end{equation}

When the near-optimal reduced-scope TMLD of (\ref{eq3-2-3}) is used, the LCIT-GSM scheme's complexity becomes
\begin{equation}\label{eq4-2-3}
    {\delta_{TMLD}} = 6M{N_r}{N_t} + 6\beta M{N_r}N,
\end{equation}
where $\frac{1}{M} \le \beta  \le 1$ is a parameter that increases with the constant $c$ of (\ref{eq3-2-2}). More specifically, we have $\beta  = \frac{1}{M}$ for $c = 1$ and $\beta  \to 1$ for $c \to \infty $, respectively. We note that as $c \to \infty $, the complexity of TMLD becomes higher than that of MLD, which is due to the fact that, for a large $c$, TMLD does not reduce the cardinality of the symbol constellation, but increases the complexity of the first step. However, in practice, a small $c$ is adequate for achieving near-ML performance at a low complexity.

When the single-stream-based DMLD of (\ref{eq3-3-1})-(\ref{eq3-3-6}) is invoked, the LCIT-GSM detection complexity encountered for a given $k$ consists of three parts: 1) the computation of ${p_k}$ in (\ref{eq3-3-1}) requires $\left( {6{N_r} + 2} \right)$ real-valued multiplications \cite{rajashekar2013reduced}; 2) $2$ and $4$ real-valued multiplications are required for obtaining ${\hat s_k}$ in PSK and QAM, respectively; 3) $6$ real-valued multiplications are required for performing (\ref{eq3-3-6}). Therefore, the LCIT-GSM detection complexity is reduced to
\begin{equation}\label{eq4-2-4}
    {\delta _{DMLD}} = \left\{ {\begin{array}{*{20}{l}}
{\left( {6{N_r} + 10} \right)N,}&{PSK}\\
{\left( {6{N_r} + 12} \right)N,}&{QAM}
\end{array}} \right..
\end{equation}
The complexity comparison of the three different detectors is summarized in Table \ref{tab3}, where the order of complexity is defined as the number of optimization metric evaluations \cite{rajashekar2013reduced}. The numerical complexity comparisons are offered in Section \ref{s5}.}}

We note that as previously shown by (\ref{eq2-1-3}), the throughput of LCIT-GSM is higher than that of GSM. This implies that for the sake of achieving the same SE, LCIT-GSM requires less TAs than its conventional GSM counterpart. As a result, although the signal detection complexity of LCIT-GSM of (\ref{eq4-2-2})-(\ref{eq4-2-4}) is generally higher than that of GSM, the proposed LCIT-GSM is still advantageous in terms of the overall energy efficiency, which may be regarded as the SE normalised by the hardware power consumption \cite{di2013spatial, xu2019sixty, basar2017index}.

\section{Simulation Results}\label{s5}
In this section, the BEP performance of LCIT-GSM systems is investigated for different numbers of TA/RAs and using different modulation schemes. For the sake of clarity, the parameters used for generating the results in this section are detailed in the figures.

\begin{figure}[!t]
	\centering
	\includegraphics[width=8.8cm]{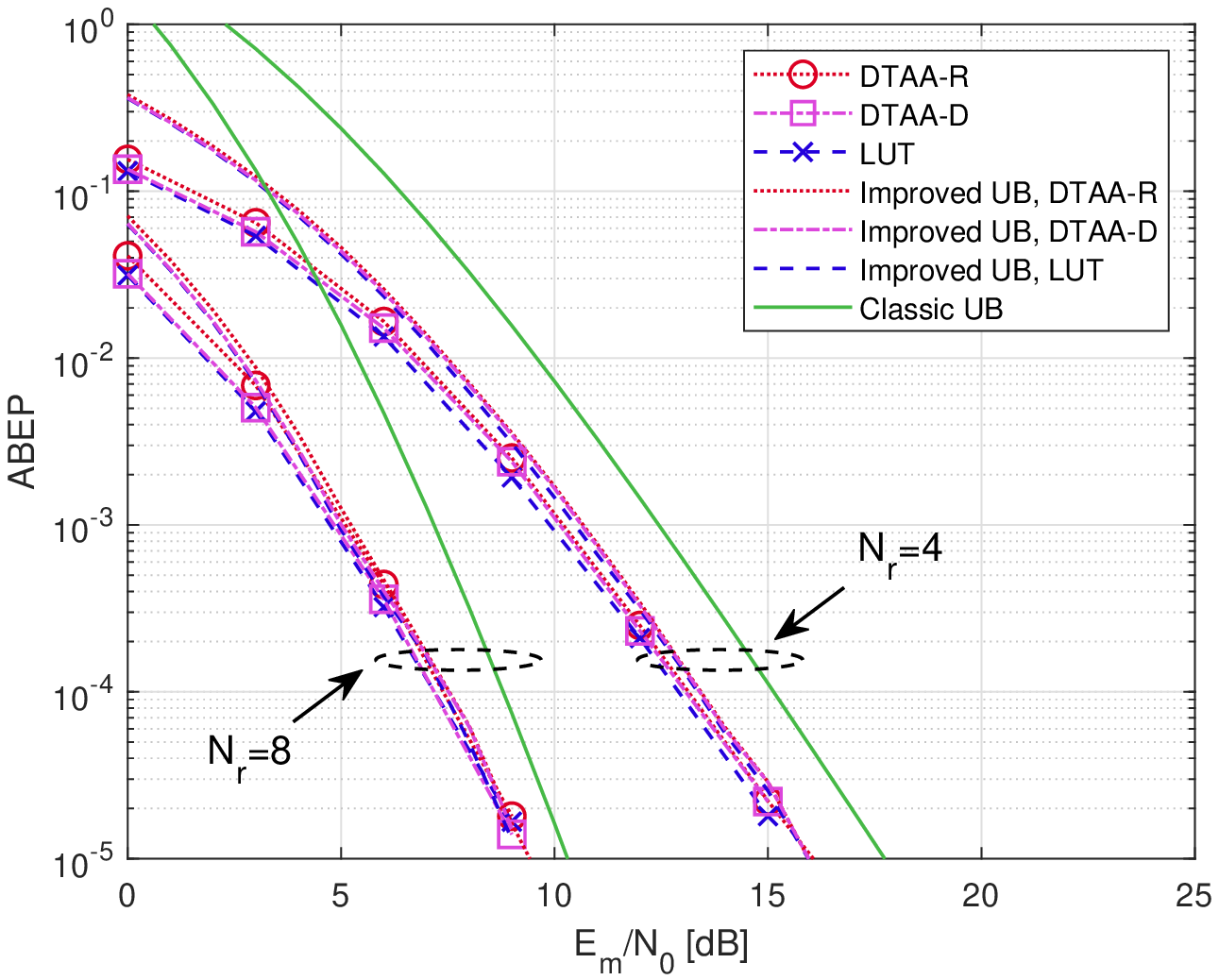}
	\caption{\color{black}{ABEP of LCIT-GSM (QPSK, $N_t=4$) versus $E_m/N_0$ under our three types of spatial mapping rules, i.e., DTAA-R in (\ref{eq2-2-1}), DTAA-D in (\ref{eq2-2-4}) and LUT in (\ref{eq2-2-5}). The classic UB and improved UB are drawn with solid line and dash-dot line, respectively. Note that LUT uses the same mapping design as conventional VGSM of \cite{osman2015variable, younis2014spatial}.}}
	\label{f5}
\end{figure}
\begin{figure}[!t]
	\centering
	\includegraphics[width=8.8cm]{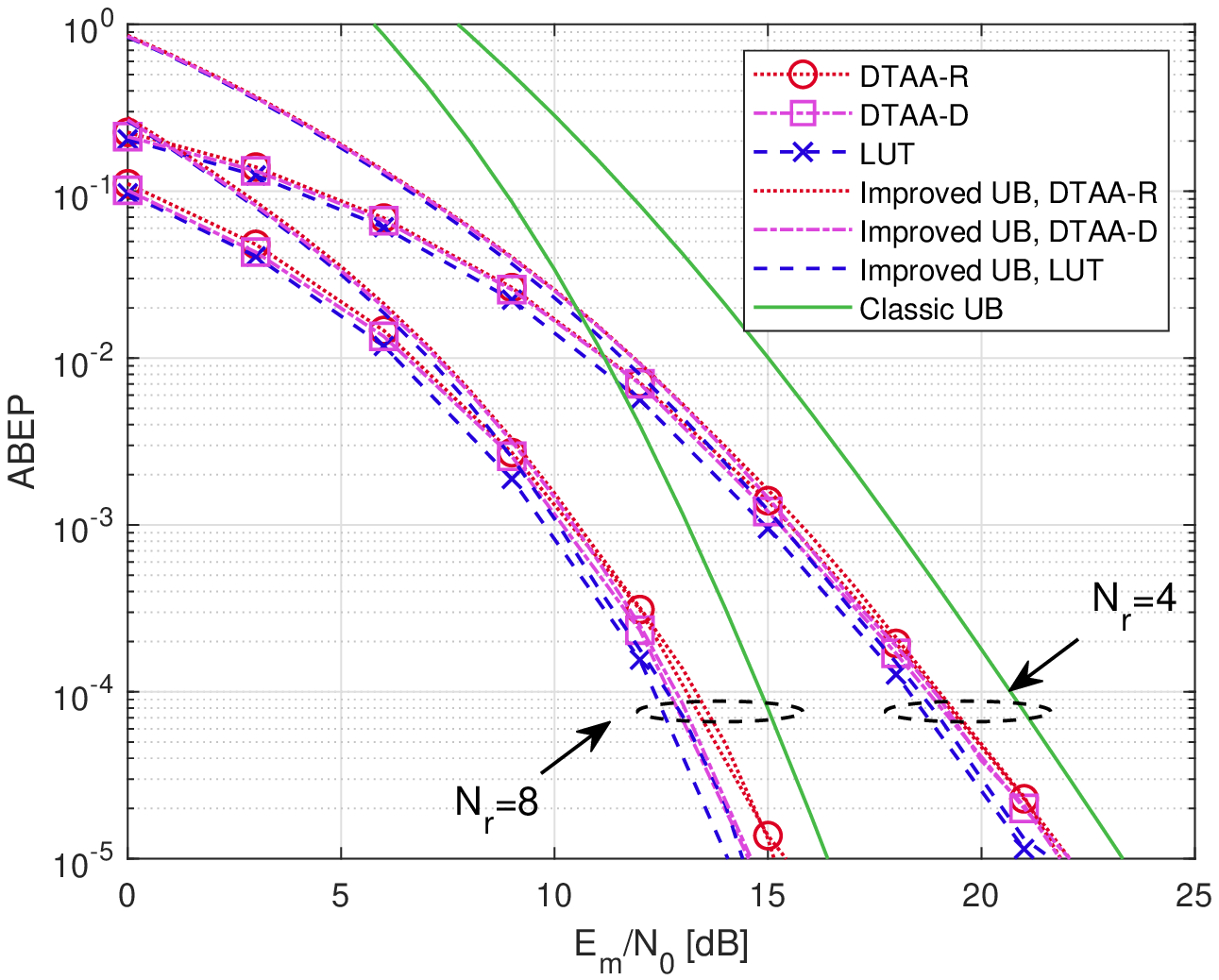}
	\caption{\color{black}{ABEP of LCIT-GSM (16QAM, $N_t=4$) versus $E_m/N_0$ under our three types of spatial mapping rules, i.e., DTAA-R in (\ref{eq2-2-1}), DTAA-D in (\ref{eq2-2-4}) and LUT in (\ref{eq2-2-5}). The classic UB and improved UB are drawn with solid line and dash-dot line, respectively. Note that LUT uses the same mapping design as conventional VGSM of \cite{osman2015variable, younis2014spatial}.}}
	\label{f6}
\end{figure}

Figs. \ref{f5} $\sim$ \ref{f8} contrast our simulations and theoretical derivations for the LCIT-GSM system in different scenarios, where the circle, square and cross patterns represent DTAA-R, DTAA-D, and LUT, respectively, while the dash-dot and solid lines represent the improved performance bound and the classic union bound, respectively. {\color{black}{The MLD is adopted in Figs. \ref{f5} $\sim$ \ref{f8} for contrasting its performance to the corresponding theoretical BEP bound.}} In Figs. \ref{f5} and \ref{f6}, the number of TAs is set to $N_{t}=4$. Although the three mapping arrangements only exhibit modest performance differences in Figs. \ref{f5} $\sim$ \ref{f8}, the DTAA-R is shown to have a worse BEP than DTAA-D and LUT, since it transmits one more bit using the same number of TAs. Furthermore, as evidenced by Figs. \ref{f5} $\sim$ \ref{f8}, the LUT performs the best. However, it is worth noting that compared to the simple DTAA-R/DTAA-D mapping, the LUT has to rely on a pre-defined LUT, which imposed a higher design complexity.

\begin{figure}[!t]
	\centering
	\includegraphics[width=8.8cm]{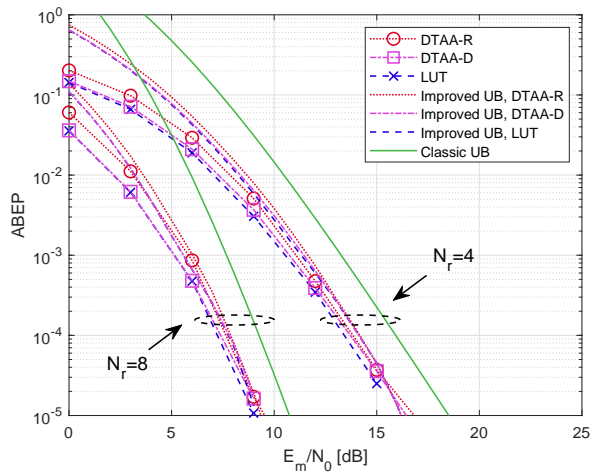}
	\caption{\color{black}{ABEP of LCIT-GSM (QPSK, $N_t=6$) versus $E_m/N_0$ under our three types of spatial mapping rules, i.e., DTAA-R in (\ref{eq2-2-1}), DTAA-D in (\ref{eq2-2-4}) and LUT in (\ref{eq2-2-5}). The classic UB and improved UB are drawn with solid line and dash-dot line, respectively. Note that LUT uses the same mapping design as conventional VGSM of \cite{osman2015variable, younis2014spatial}.}}
	\label{f7}
\end{figure}
\begin{figure}[!t]
	\centering
	\includegraphics[width=8.8cm]{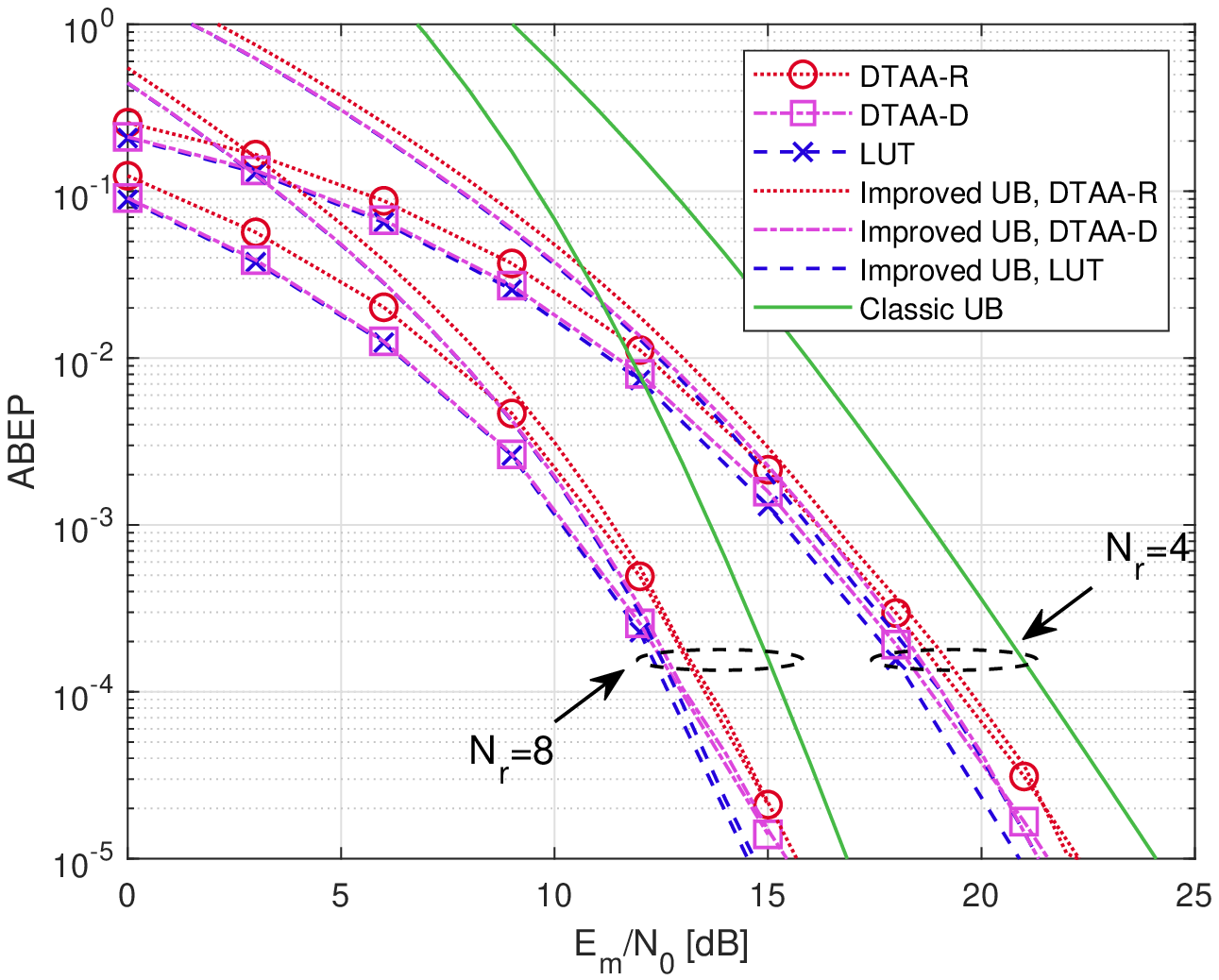}
	\caption{\color{black}{ABEP of LCIT-GSM (16QAM, $N_t=6$) versus $E_m/N_0$ under our three types of spatial mapping rules, i.e., DTAA-R in (\ref{eq2-2-1}), DTAA-D in (\ref{eq2-2-4}) and LUT in (\ref{eq2-2-5}). The classic UB and improved UB are drawn with solid line and dash-dot line, respectively. Note that LUT uses the same mapping design as conventional VGSM of \cite{osman2015variable, younis2014spatial}.}}
	\label{f8}
\end{figure}

\begin{figure}[!t]
	\centering
	\includegraphics[width=8.8cm]{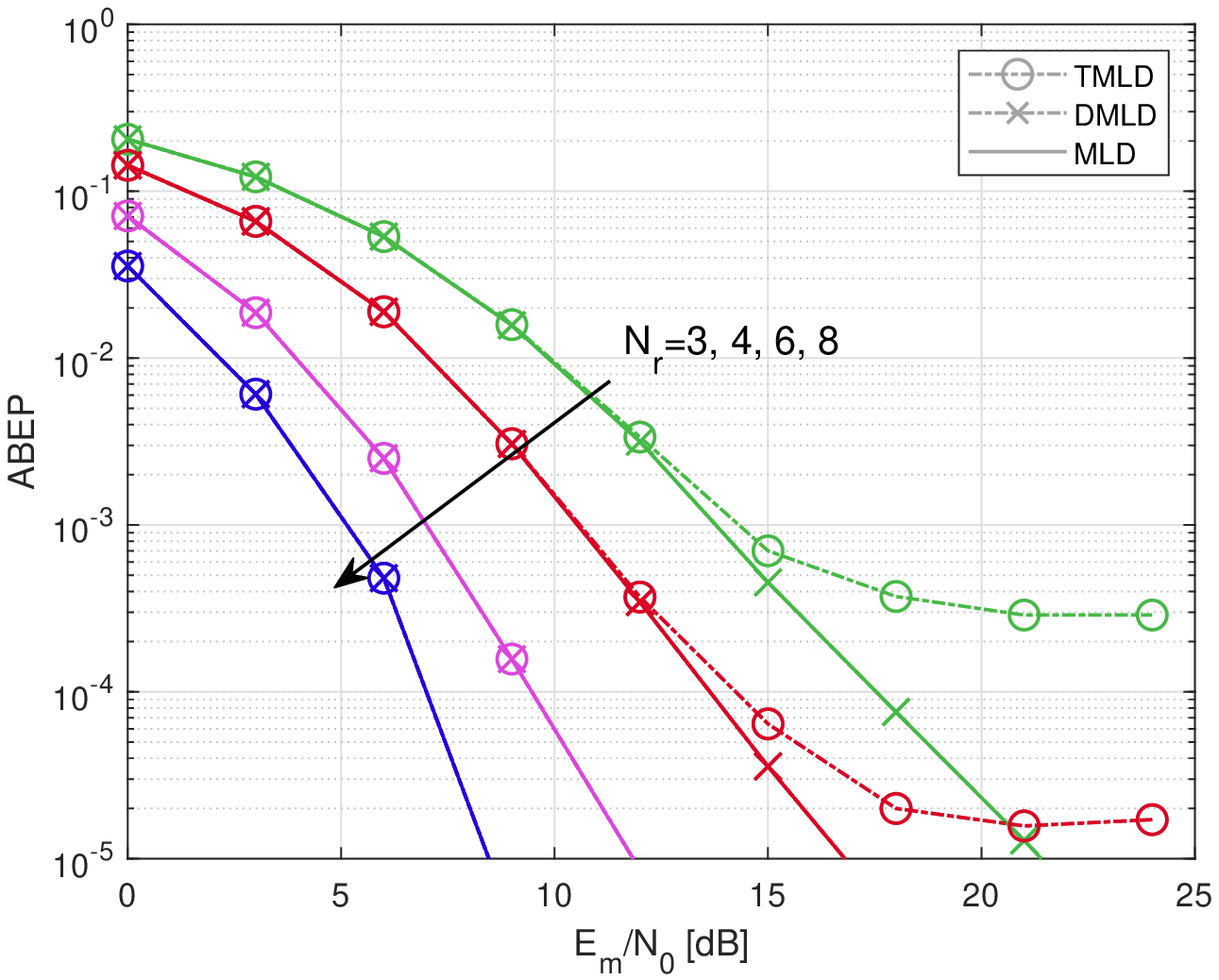}
	\caption{\color{black}{ABEP of LCIT-GSM (QPSK, $N_t = 6$) versus $E_m/N_0$ for our three types of detectors. The LUT is used to illustrate the compatibility of the proposed low-complexity detectors with conventional VGSM of \cite{osman2015variable, younis2014spatial}.}}
	\label{f9}
\end{figure}
\begin{figure}[!t]
	\centering
	\includegraphics[width=8.8cm]{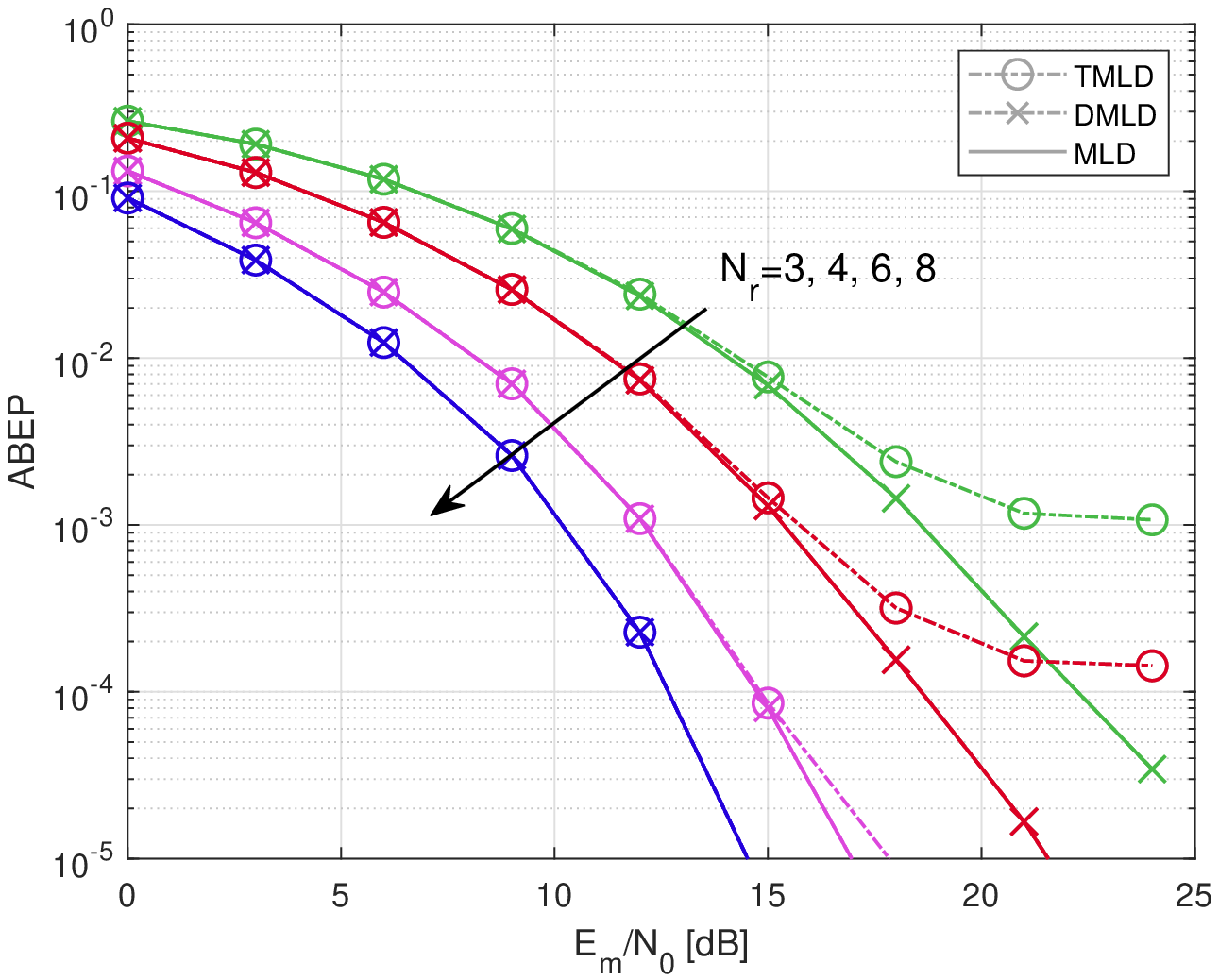}
	\caption{\color{black}{ABEP of LCIT-GSM (16QAM, $N_t = 6$) versus $E_m/N_0$ for our three types of detectors. The LUT is used to illustrate the compatibility of the proposed low-complexity detectors with conventional VGSM of \cite{osman2015variable, younis2014spatial}.}}
	\label{f10}
\end{figure}
Moreover, it is demonstrated by Figs. \ref{f5} $\sim$ \ref{f8} that the improved theoretical bound is about 1.5dB tighter than the conventional bound. Figs. \ref{f7} and \ref{f8} depict the theoretical derivation and simulation results of using ${N_t} = 6$ and QPSK/16QAM signalling. It can be seen in Figs. \ref{f7} and \ref{f8} that, once again, the improved performance bound has a 1.5dB performance improvement over the conventional performance bound. Furthermore, as the number of RAs increases, the performance advantage of LUT over DTAA-R/DTAA-D improves, and the UB also becomes tighter in Figs. \ref{f7} and \ref{f8}, which once again verifies the reliability of our conclusion.

\begin{figure}[!t]
	\centering
	\includegraphics[width=8.8cm]{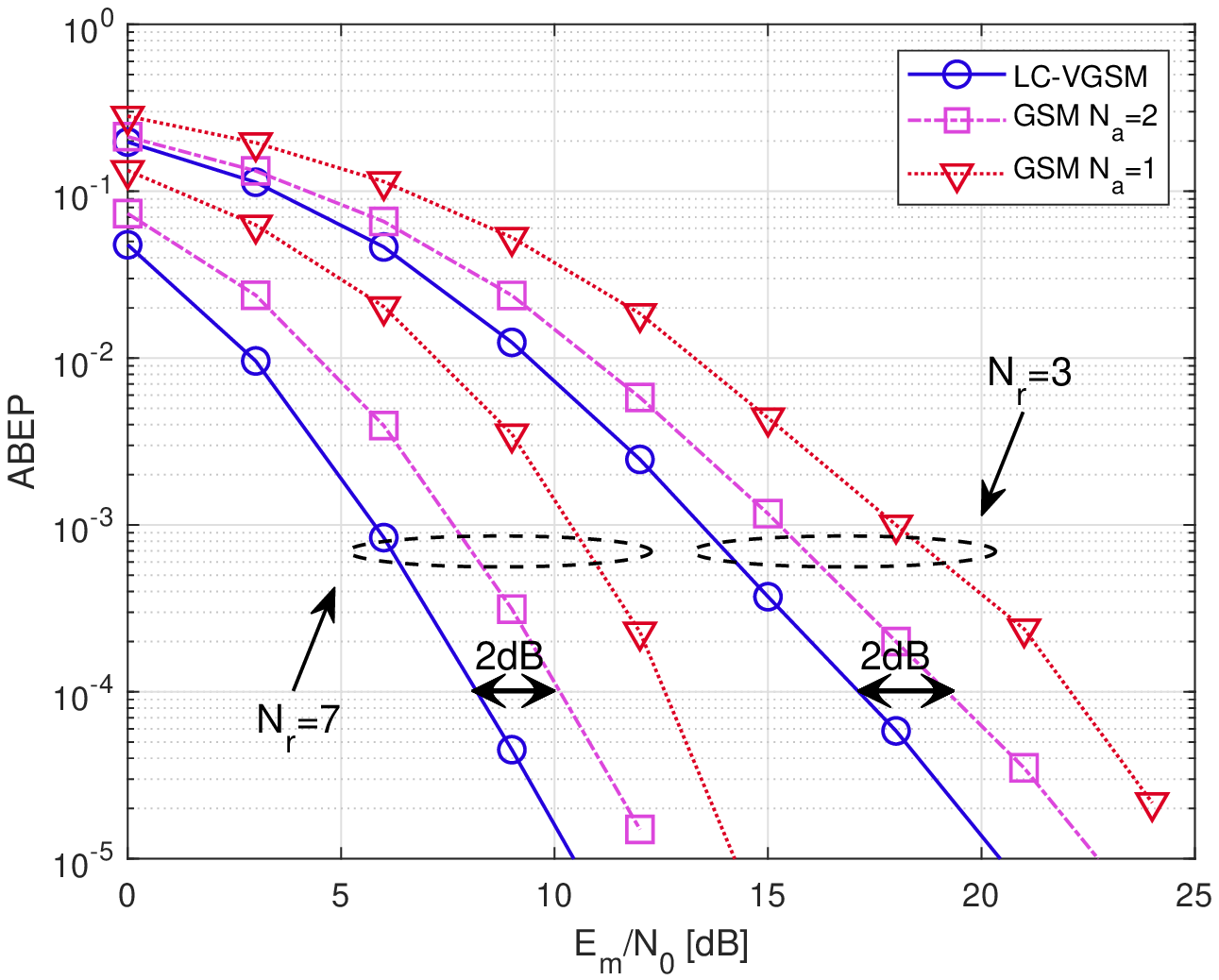}
	\caption{The performance comparison between LCIT-GSM (QPSK) and GSM for different ${N_a }$. The LUT-LCIT-GSM with $N_t = 5$ is used to maintain a overall rate of 6bpcu.}
	\label{f11}
\end{figure}
\begin{figure}[!t]
	\centering
	\includegraphics[width=8.8cm]{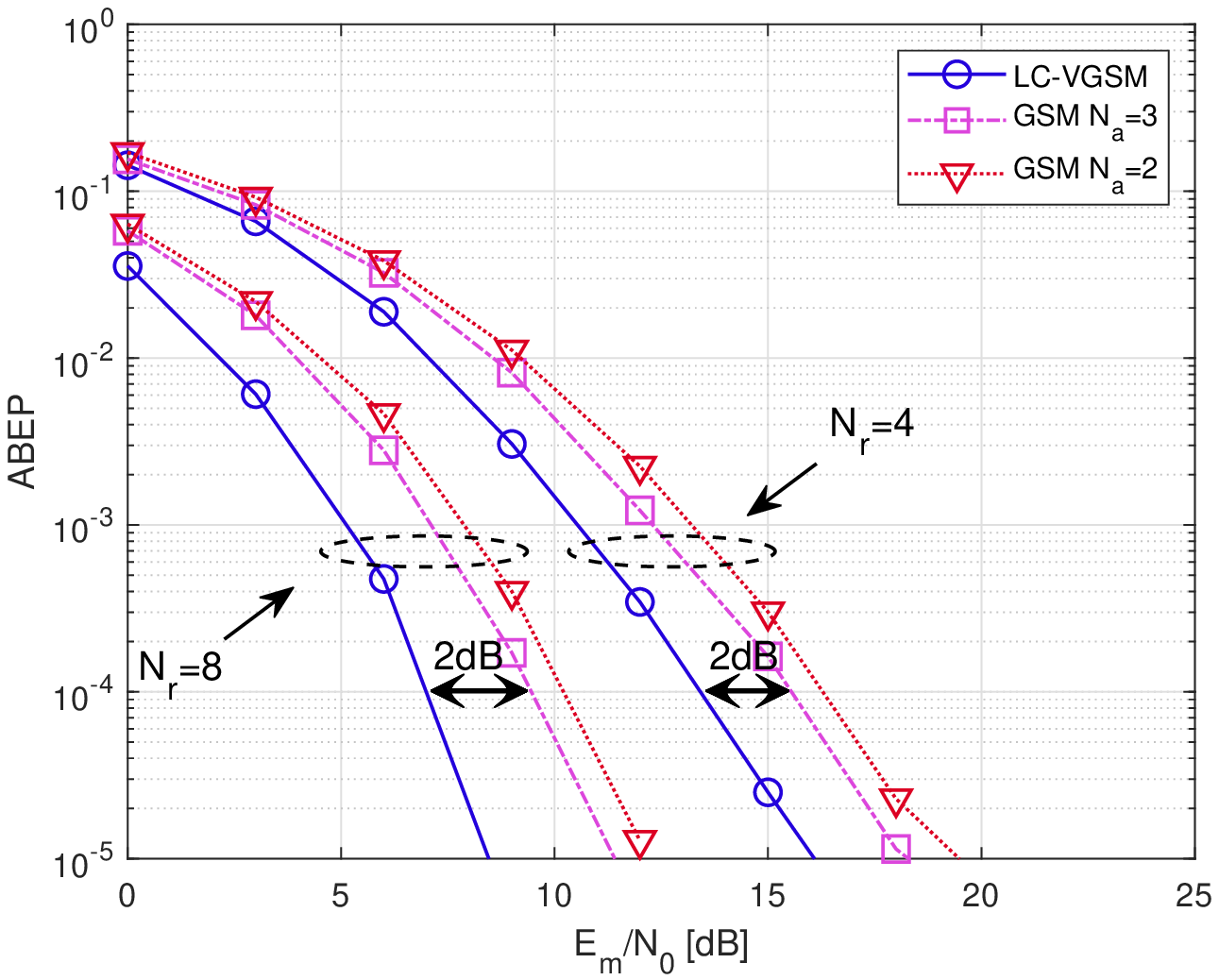}
	\caption{The performance comparison between LCIT-GSM (QPSK) and GSM for different ${N_a }$. The LUT-LCIT-GSM with $N_t = 6$ is used to maintain a overall rate of 7bpcu.}
	\label{f12}
\end{figure}

Figs. \ref{f9} and \ref{f10} compare the performance of different detectors for our LCIT-GSM system having ${N_t} = 6$. As discussed in Section \ref{s4-2}, the TMLD exhibits a lower detection complexity, but it also imposes a performance loss compared to the MLD and DMLD, as confirmed by Figs. \ref{f9} and \ref{f10}. We note that the constant $c$ of (\ref{eq3-2-2}) is set to 1.5 in Figs. \ref{f9} and \ref{f10}, which may be increased for the sake of improving the performance, but this would also result in a higher detection complexity according to (\ref{eq4-2-3}). Moreover, Figs. \ref{f9} and \ref{f10} show that although TMLD performs similarly to MLD at low SNRs, the TMLD may exhibit an error floor. {\color{black}{The error floor will move down upon increasing ${N_r}$ because the channels between different TAs tend to become more orthogonal for an increased number of RAs. More specifically, the index of the modulated symbol found in the first step becomes more accurate and hence supports the global optimality of the second step of local search. Fortunately, DMLD always maintains the same performance as MLD, which benefits from the fact that for a regular constellation, the rounding-based detector and the exhaustive search return the same result.}} The simulation results of 16QAM associated with ${N_t} = 6$ are presented in Fig. \ref{f10}, which leads to the same observations on TMLD as Fig. \ref{f10}. Moreover, by comparing the performance of QPSK and 16QAM in Figs. \ref{f9} and \ref{f10}, respectively, we can see that the performance of QPSK is always better than that of 16QAM.

\begin{figure}[!t]
	\centering
	\includegraphics[width=8.8cm]{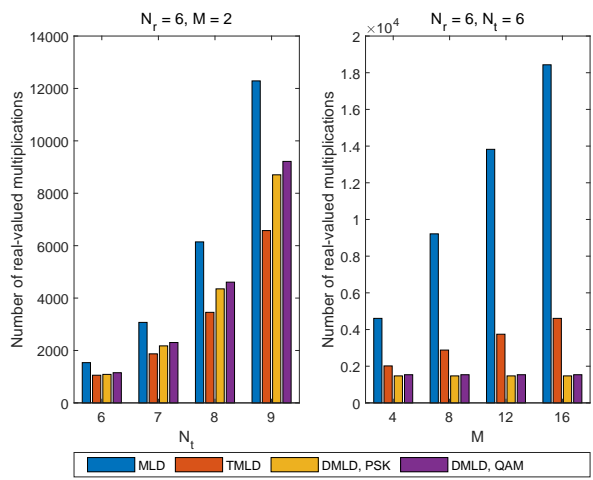}
	\caption{\color{black}{Complexity comparison of different detectors for LCIT-GSM with ${N_r} = 6$, where DTAA-D and LUT are adopted.}}
	\label{f13}
\end{figure}
The BEP performance comparisons between GSM and LCIT-GSM are offered in Figs. \ref{f11} and \ref{f12}. In Fig. \ref{f11}, the number of TAs is set to $N_{t} = 5$ and QPSK is used for LCIT-GSM. In order to maintain the overall throughput of 6bpcu, 16QAM and 8QAM are used for GSM with the fixed number of active TAs ${N_a } = 1$ and ${N_a } = 2$, respectively. It can be seen in Fig. \ref{f11} that LCIT-GSM outperforms GSM for ${N_a } = 2$ by about 2dB for different numbers of RAs of $N_{r} = 3$ and $N_{r} = 7$. When GSM associated with ${N_a } = 1$ is considered, a 4dB performance gain can be observed in Fig. \ref{f11}. In Fig. \ref{f12}, our performance comparisons between GSM and LCIT-GSM are portrayed for the case of $N_{t} = 6$. The QPSK constellation is also adopted for LCIT-GSM, while 16QAM and 8QAM are used for GSM at ${N_a } = 2$ and ${N_a } = 3$, respectively, hence maintaining a overall rate of 7bpcu. The 2dB performance gain compared to GSM using ${N_a } = 3$ can also be observed in Fig. \ref{f12} to verify the advantage of the proposed LCIT-GSM scheme.

{\color{black}{Finally, in Fig. \ref{f13}, the computational complexity of different detectors is compared in terms of the number of real-valued multiplications, where we set $c = 1$ and when the DTAA-D and LUT are adopted. As seen in Section \ref{s4-2}, the complexity of MLD is prohibitive and increases exponentially with ${N_t}$, while our proposed detectors achieve lower complexity for all the scenarios considered. More specifically, the complexity of the TMLD is lower than that of the DMLD for a small constellation, and the complexity-difference will widen as the number of TAs increases. This is because the ratio of the DMLD's complexity to $N$ is higher than that of TMLD, bearing in mind that $N = {2^{{N_t} - 1}}$ for DTAA-D and LUT. By contrast, for a moderate number of TAs, the DMLD is more competitive in conjunction with high-order modulation, which is due to the fact that the complexity of the DMLD does not increase with the constellation size $M$. In summary, the proposed low-complexity detectors strike an attractive BEP performance vs. detection complexity tradeoff and can be applied flexibly in various scenarios. Moreover, we emphasize that the DMLD is always optimal but it is only suitable for regular constellations, while the TMLD is suboptimal, but it can strike a flexible BEP performance vs. computational complexity trade-off by adjusting the constant $c$ in (\ref{eq3-2-2}).}}

\section{Conclusions}\label{s6}
{\color{black}{A novel LCIT-GSM scheme was proposed, which is capable of maximising the GSM throughput without imposing IAI while reducing the system complexity of the conventional VGSM of \cite{osman2015variable, younis2014spatial}.}} In contrast to the conventional GSM that always invokes a fixed number $N_{a}$ of activated TAs, the proposed LCIT-GSM utilises all legitimate TA activation patterns, where $N_{a}$ becomes a variable that has to be recovered by the receiver. Nonetheless, in the absence of IAI, a low-complexity single-stream based detector may be invoked by the LCIT-GSM receiver. Both our theoretical derivation and our numerical simulation have confirmed that LCIT-GSM increases the overall SE logarithmically with the number of TAs. More specifically, for the LCIT-GSM mapping, the arrangements of DTAA-R, DTAA-D and LUT are conceived in order to facilitate a flexible MIMO deployment with special focus on the all-zero spatial beam stream. Furthermore, as for the LCIT-GSM signal detection, we have developed the low-complexity TMLD and DMLD schemes, where the former adopts the sphere decoding philosophy for the sake of striking a compelling performance vs complexity trade-off, while the latter is capable of achieving the ML performance at a single-stream detection complexity that does not grow with the number of modulation levels. Furthermore, an improved theoretical UB is derived for the proposed LCIT-GSM, which is confirmed by our Monte-Carlo-based simulation to be about 1.5dB tighter than the existing conventional solution. Finally, our simulation results demonstrate that the proposed LCIT-GSM scheme achieves improved performance over the conventional GSM and lower complexity than the VGSM. {\color{black}{In this paper, three spatial mapping schemes are presented for exploiting the low complexity of the bitwise TA activation. We believe that the family of mapping schemes designed by formal optimization methods may achieve a better BEP at the cost of a moderate design complexity, which is left as a topic for our future research.}}

\bibliographystyle{IEEEtran}
\bibliography{An}
\end{document}